\documentclass[reprint,superscriptaddress,a4paper,aps,showpacs,longbibliography]{revtex4-1}
\usepackage{graphicx}
\graphicspath{{./}}
\usepackage{amsfonts}
\usepackage{amsbsy}
\usepackage{amsmath}
\usepackage{diagbox}

\begin{document}

\title{Dual nature of magnetism in MnSi}

\author{A. Yaouanc}
\affiliation{Universit\'e Grenoble Alpes, CEA, IRIG-PHELIQS, F-38000 Grenoble, France}
\author{P. Dalmas de R\'eotier}
\affiliation{Universit\'e Grenoble Alpes, CEA, IRIG-PHELIQS, F-38000 Grenoble, France}
\author{B. Roessli}
\affiliation{Laboratory for Neutron Scattering and Imaging, Paul Scherrer Institute, CH-5232 Villigen-PSI, Switzerland}
\author{A. Maisuradze}
\affiliation{Department of Physics, Tbilisi State University, Chavchavadze 3, 
GE-0128 Tbilisi, Georgia}
\author{A. Amato}
\affiliation{Laboratory for Muon-Spin Spectroscopy, Paul Scherrer Institute,
CH-5232 Villigen-PSI, Switzerland}
\author{D. Andreica}
\affiliation{Faculty of Physics, Babes-Bolyai University, 400084 Cluj-Napoca, Romania}
\author{G. Lapertot}
\affiliation{Universit\'e Grenoble Alpes, CEA, IRIG-PHELIQS, F-38000 Grenoble, France}

\date{\today}

\begin{abstract}

The temperature dependence of the manganese magnetic moment and the spin-lattice relaxation rate measured by the muon spin relaxation technique in the magnetically ordered phase of the chiral intermetallic cubic MnSi system are both explained in terms of helimagnon excitations of a localized spin model. The two free parameters characterizing the helimagnon dispersion relation are determined.  A combined analysis of the two data sets cannot be achieved using the self-consistent renormalization theory of spin fluctuations which assumes the magnetism of MnSi to arise uniquely from electronic bands.  As a result of this work, MnSi is proposed to be a dual electronic system composed of localized and itinerant magnetic electrons. Finally we note that the analysis framework can be applied to other helimagnets such as the magnetoelectric compound Cu$_2$OSeO$_3$.

\end{abstract}


\maketitle
 
\section{Introduction}
\label{Intro}

The history of the metallic compound MnSi is quite rich and spans almost 90 years.  Its crystal structure was established in 1933 \cite{Boren33}. It crystallizes into the cubic $P2_13$ space group which lacks a center of symmetry, giving rise to the possibility of Dzyaloshinski-Moriya (DM) exchange interactions \cite{Dzyaloshinskii58,Moriya60}. Its magnetic structure below its magnetic ordering temperature $T_c \approx 29.5$~K was found to be characterized by a propagation wavevector ${\bf k}$ parallel to a $\langle 111\rangle$ crystal direction and with a modulus $k \simeq 0.35 \, {\rm nm}^{-1}$ at low temperature \cite{Ishikawa76}. A field-theory model was built to explain it \cite{Bak80,Nakanishi80}. MnSi served as a model system for the development of the self-consistent renormalization (SCR) theory. This theory for spin fluctuations in itinerant electron systems quantitatively interprets the paramagnetic susceptibility and the value of $T_c$ \cite{Lonzarich85,Moriya85,Kakehashi13}. 

Since the advent of the new century, unabated new key results have appeared. Among them, an extended non-Fermi-liquid regime in the paramagnetic phase under pressure was revealed \cite{Pfleiderer01} and its association with an exotic diffuse magnetic neutron scattering was discovered \cite{Pfleiderer04}, although this latter result has recently been disputed \cite{Bannenberg19}. A small pocket in the field-temperature phase diagram was shown to host a magnetic skyrmion lattice \cite{Muhlbauer09a}. The magnetic structure at low temperatures was recently refined from zero-field data recorded with the muon spin rotation/relaxation ($\mu$SR) technique  \cite{Dalmas16}. It deviates from the originally purported pure helical magnetic structure; see also Refs.~\onlinecite{Dalmas17,Dalmas18}. 

\begin{figure}
  \begin{picture}(245,220)
    \put(0,102){\includegraphics[width=\linewidth]{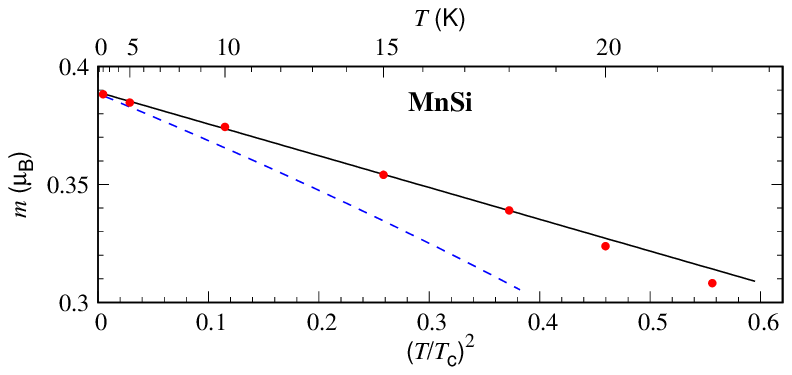}}
    \put(35,154){\small (a)}
    \put(0,0){\includegraphics[width=\linewidth]{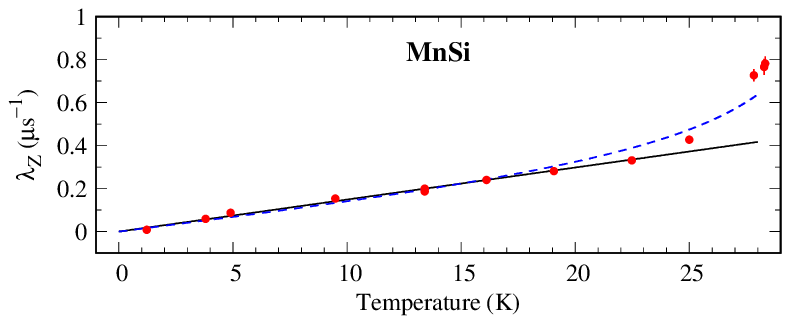}}
    \put(35,55){\small (b)}
   \end{picture}
\caption{
Temperature dependence of two parameters characterizing MnSi, extracted from $\mu$SR data. The experimental uncertainties are smaller than the bullet size except for $\lambda_Z$ above 27~K. (a) $m$ versus $(T/T_c)^2$. The solid line results from a fit of $m(T) = m(0) [1 - a(T/T_{\rm c})^2 ]$ to the data up to 18~K, where $m(T=0) = 0.3891 \, (5) \, \mu_{\rm B}$ --- in agreement with a previous estimate \cite{Thessieu98a} --- and $a = 0.346\,(6)$~K. In addition, $T_c = 29.5$~K. (b) $\lambda_Z$ versus temperature from Ref.~\cite{Yaouanc05} with the solid line resulting from a linear fit up to 22.5~K with a slope $b_\lambda = 14.9 \, (1) $\,ms$^{-1}$K$^{-1}$. The dashed lines in both panels are computed from the SCR theory \cite{Lonzarich85,Moriya74,Yaouanc11}: $m(T) = m(T = 0) \left [1 - (T/T_c )^2 \right ]^{1/2}$ and $\lambda_Z(T) = c_\lambda T/m(T)$ where $c_\lambda$ = 5270\,(170)~$\mu_{\rm B}$\,K$^{-1}$\,s$^{-1}$ results from a fit, i.e.\ it is not SRC predicted.
}
\label{experimental_data}
\end{figure}

This report provides a consistent interpretation based on helimagnon excitations of the quadratic and linear temperature dependencies, at low temperature, of the manganese magnetic moment $m$ and the muon spin-lattice relaxation rate $\lambda_Z$, respectively; see Fig.~\ref{experimental_data} for the data and fits. 
The SCR theory is unable to account for $m(T)$ as shown in Fig.~\ref{experimental_data}(a). As a result of our analysis, we propose to view MnSi as a dual system where an electronic subset of itinerant states coexists with a subset of localized electrons, the helimagnons being a signature of the latter electrons. 

The organization of the paper is as follows.  Section~\ref{Structure} focuses on the refinement of the magnetic structure of MnSi up to 27.5~K based on zero-field $\mu$SR spectra. In Sec.~\ref{helimagnons} we recall the helimagnon excitation picture. Section \ref{helimagnons_T} provides a description of $m(T)$. The following section (Sec.~\ref{helimagnons_rate}) deals with $\lambda_Z(T)$. A discussion of the results obtained in this work is given Sec.~\ref{discussion}. It is followed by conclusions and comments in Sec.~\ref{conclusion}. The theoretical modelling of the relaxing $\mu$SR polarization function is described in Appendix~\ref{Long_pol}. Appendix~\ref{Aux} outlines miscellaneous results used in Appendix~\ref{Long_pol}.

\section{Determination of the zero-field magnetic structure}
\label{Structure}

We first expose information on MnSi that is useful for the analysis of the $\mu$SR spectra. Then we provide experimental details and a discussion of the  experimental results leading to the determination of the zero-field magnetic structure up to 27.5~K.

\subsection{Some structural information}
\label{Structure_info}

We specify the position of a unit cell by the cubic lattice vector $ {\bf i }$ and a manganese atom within a cell by ${\bf d}$. Four such ${\bf d}$ vectors exist because of the four manganese atoms in the unit cell. The magnetic moment at position $ {\bf i + d}$ is written as ${\bf m}_{i + d}$ with 
\begin{eqnarray}
\frac{{\bf m}_{i + d}}{m}
& = & \cos \left[ {\bf k}\cdot ({\bf i} + {\bf d})\right]{\bf a}_d - \sin \left[{\bf k}\cdot ({\bf i}+ {\bf d})\right]{\bf b}_{d},
 \label{helix_general}
\end{eqnarray}
where the vectors $({ {\bf a}}_{d}, { {\bf b}}_{d}, {\bf k}/k )$ form a direct orthonormal basis. Because of the symmetry imposed by the ${\bf k}$ direction, the four ${\bf m}_{i + d}$ split into two families, i.e.\ the so-called orbits \cite{Dalmas16}. One orbit is constituted by the manganese site for which the local three-fold symmetry axis is collinear to ${\bf k}$. Conversely at the three other sites the local three-fold axis is not collinear to ${\bf k}$. These three sites constitute the second orbit. In fact, the magnetic moments at a given site form a regular helical structure, but with an utmost peculiarity: while the third Euler angle characterizing the orientation of $({ {\bf a}}, { {\bf b}}, {\bf k}/k)$ in the cubic crystallographic frame is set --- without loss of generality --- to zero for the first manganese site, it is a finite free parameter $\psi_{111}$ for the other three sites. If the magnetic structure were helical, we would have $\psi_{111} =0$. We have reported $\psi_{111} = -2.04 \, (11)$~degrees at 5~K for a Czochralski crystal and $\psi_{111} = -2.11 \, (11)$~degrees for a Zn-flux sample at the same temperature \cite{Dalmas18}, i.e.\ equal values within experimental uncertainties. This value might seem negligible. However, this phase shift is appreciable when compared to the variation of the phase ${\bf k}\cdot({\bf i} + {\bf d})$ between adjacent [111] manganese planes. It varies in increments of 2.91 or 2.36 degrees.

\subsection{Experimental}
\label{Structure_exp}

The zero-field (ZF) experiments were performed using a single crystal prepared by Czochralski pulling from a stoichiometric melt, and cut in the form of a platelet perpendicular to the crystallographic [111] direction. The measurements were carried out in zero-field at the general purpose surface-muon instrument (GPS) of the Swiss Muon Source located at the Paul Scherrer Institute (Switzerland); see Ref.~\onlinecite{Amato17} for a recent description of the spectrometer. They give access to the so-called asymmetry from which the field distribution at the muon sites $D_{\rm osc} (B)$ is extracted \cite{Yaouanc11}. This distribution arises from the incommensurate nature of the magnetic structure. The local field at a muon site is the sum of the dipole fields generated by the magnetic moments at the manganese atoms and the contact field reflecting the conduction electron polarization at the muon sites. Owing to the dipole field anisotropy, details of the magnetic structure can be unravelled.

\subsection{Experimental results}
\label{Structure_results}

\begin{figure}[t]
{\includegraphics[width=\linewidth ]{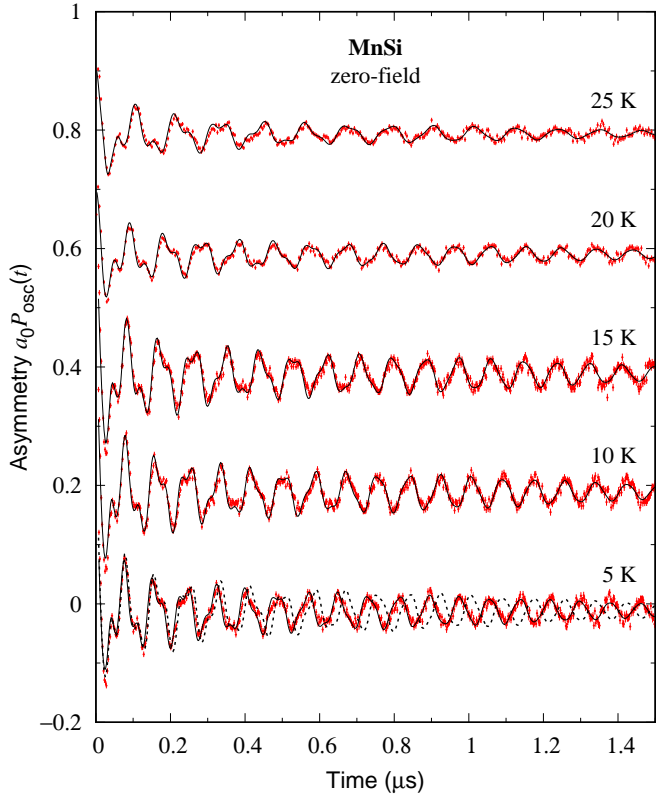}}
\caption{
Precessing contribution to ZF $\mu$SR asymmetry spectra recorded at five temperatures spanning the temperature range from 5 to 25~K. The solid lines result from fits as described in the main text. The dotted line shown together with the 5~K spectrum is a duplicate of the result of the fit to the data recorded at 10~K. It illustrates the extreme sensitivity of the spectra to the value of $m$ which decreases by a mere 2.7\% between 5 and 10~K. The data recorded at 10~K and above are vertically shifted for a better vision. 
}
\label{Fig_spectra}
\end{figure}

\begin{figure}[t]
{\includegraphics[width=\linewidth ]{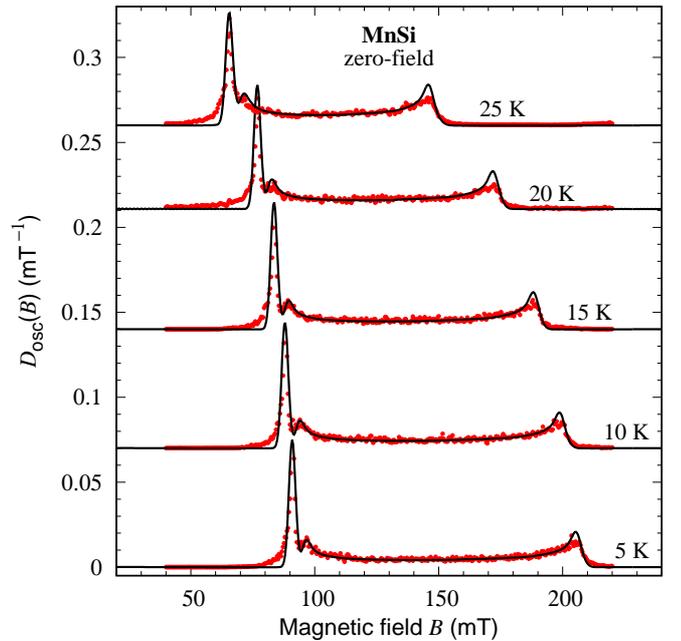}}
\caption{
Magnetic field distributions $D_{\rm osc}(B)$ associated with the data displayed in Fig.~\ref{Fig_spectra}. The solid lines result from the fits shown in Fig.~\ref{Fig_spectra}, i.e.\ the $D_{\rm osc}(B)$ data are not fitted. The distributions at 10~K and above are vertically shifted for a better vision. 
}
\label{Fig_distributions}
\end{figure}

Examples of asymmetry spectra are displayed in Fig.~\ref{Fig_spectra}. The data were recorded over a period spanning several years in which different experimental conditions were used, e.g.\ the orientation of the initial muon polarization relative to the muon momentum, which explains the difference in the amplitude of the oscillations. Figure~\ref{Fig_distributions} displays the distributions corresponding to the asymmetry spectra. Such distributions are computed from the measured asymmetry spectra using the reverse Monte Carlo technique supplemented with the maximum entropy requirement \cite{Yaouanc17,Maisuradze18}. The overall shape of the distributions remains similar for all the distributions while they are clearly shifted towards low field when the temperature increases: this is a direct measurement of the temperature dependence of $m$.

The solid lines in Fig.~\ref{Fig_spectra} are the result of fits to the model exposed in Ref.~\cite{Dalmas16}. Parameters of different origin enter the model. The muon site --- given by the muon reduced coordinate $x_\mu = 0.532$ --- and the coupling of its spin with the conduction electron spins --- characterized by parameter $r_\mu H/4 \pi = -1.04$ --- are fully consistent with earlier measurements in a large magnetic field in the paramagnetic phase \cite{Amato14}; see also Refs.~\onlinecite{Dalmas17,Dalmas18}. The modulus of the magnetic structure propagation wavevector \footnote{The modulus of the propagation wavevector increases from $k \simeq 0.35 \, {\rm nm}^{-1}$ for $T \rightarrow 0$ to $k \simeq 0.38 \, {\rm nm}^{-1}$ for $T \rightarrow T_c$ \cite{Ishikawa76,Grigoriev06,Janoschek13}. This variation has a negligible influence on $D_{\rm osc}$; see Fig.~1(a) of Ref.~\onlinecite{Dalmas18}} is taken from small angle neutron scattering measurements. The sensitivity of a distribution to the different physical parameters is illustrated in Ref.~\onlinecite{Dalmas18}.

\begin{figure}
{\includegraphics[width=\linewidth]{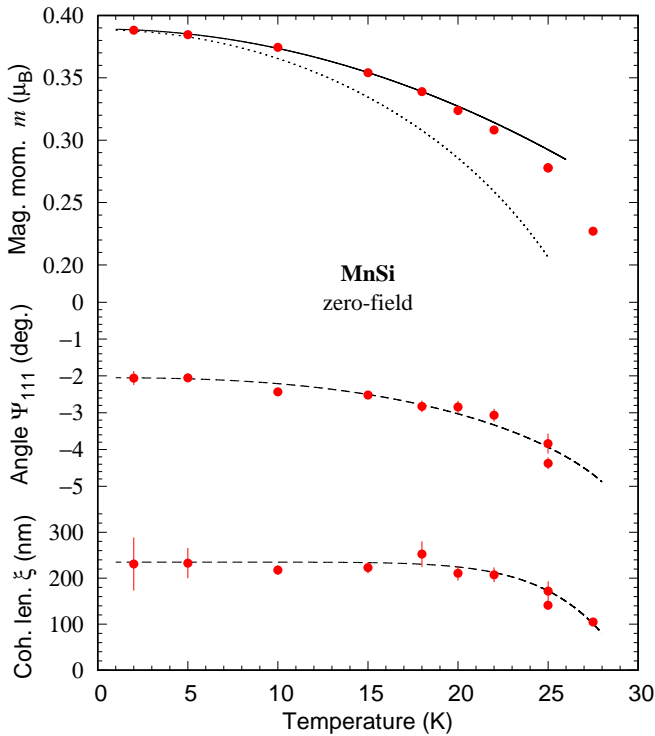}}
\caption{
Temperature dependence of three parameters related to the ZF magnetic structure of MnSi: the amplitude of the magnetic moment per manganese atom $m$, the phase shift between the two orbits $\psi_{111}$, and the magnetic structure coherence length $\xi$. The solid line in the upper panel results from the helimagnon model as explained in the caption of Fig.~\ref{experimental_data}. The dotted line is the prediction of the SCR theory \cite{Lonzarich85}: $m(T) = m(0)[1-(T/T_{\rm c})^2]^{1/2}$. While the experimental uncertainties are always drawn, they are smaller than the bullet size for the magnetic moment. The dashed lines in the lower two panels are guides to the eye.}
\label{parameters_structure}
\end{figure}

Figure~\ref{parameters_structure} displays the temperature dependence of three refined parameters: the magnetic moment $m$, the phase shift $\psi_{111}$, and the magnetic structure coherence length $\xi$. The determination of $m$ is remarkably accurate.
Two measurements have been performed at 25~K in different campaigns of measurements. The values cannot be distinguished in the $m(T)$ panel while the higher statistics of the second measurement is reflected in the smaller uncertainties for $\psi_{111}$ and $\xi$. 

We acknowledge that the satellite peak in $D_{\rm osc}(B)$ at higher temperatures becomes less resolved whereas the model appears to show no change. However, the data of Fig.~\ref{parameters_structure} are extracted from fits to the asymmetry spectra which are well described in the whole temperature range; see Fig.~\ref{Fig_spectra}. A field distribution plot is only a practical tool for a first qualitative understanding of a spectrum. 

The value of $\xi$ at low temperature is consistent with the SANS lower bound of 200~nm \cite{Pfleiderer04}. Its decrease while approaching $T_{\rm c}$ is conventional. Much less conventional is the temperature dependence of $\psi_{111}$. In absolute value it increases by a factor 2 between 2 and 25~K. This behavior will be discussed in a separate paper.

\section{The helimagnon picture}
\label{helimagnons}

We first write the simplest possible magnetic energy expression and then provide the helimagnon dispersion relation deduced from it.

\subsection{Setting the stage}
\label{helimagnons_stage}

We shall use three  orthonormal reference frames. Since the magnetic structure is characterized by a finite ${\bf k}$, it is natural to introduce the frame $({\bf a}, {\bf b}, {\bf n})$, with ${\bf n} = {\bf k}/k$. We shall find sufficient the description of the helimagnon excitations in the field-theory framework of Ref.~\onlinecite{Bak80}. So we do not need to distinguish the four Mn magnetic moments in the unit cell, i.e.\ we write ${\bf a}$ and ${\bf b}$ rather than ${\bf a}_d$ and ${\bf b}_d$. The magnetic energy operator ${\mathcal E}$ --- in fact we will treat it as a Hamiltonian, see Appendix \ref{Hamiltonian} --- is easily diagonalized in the local coordinate frame $({\bf x}, {\bf y}, {\bf z})$ rotating with the helical structure \cite{Nagamiya67}. We take the unit vector ${\bf z}$ parallel to the magnetic moment, add a second vector ${\bf y} = {\bf n}$ and complete the set with ${\bf x} = {\bf y}\times {\bf z}$. Finally, we have the laboratory reference frame $({\bf X}, {\bf Y}, {\bf Z})$ with ${\bf Z}$ parallel to the initial muon beam polarization taken along the $[111]$ crystal axis \cite{Dalmas97,Dalmas04,Yaouanc11}. 

 We write down ${\mathcal E}$ as the sum of Heisenberg and Dzyaloshinski-Moriya (DM) exchange energies with $B_1$ and $D$ measuring their respective strength \cite{Bak80}. In the small wavevector limit of interest for MnSi,
\begin{eqnarray}
{\mathcal E} 
&  = & V \int \left [\frac{B_1 q^2}{2} |{\bf S}_{\bf q}|^2  
  + i D {\bf q} \cdot ({{\bf S}}_{\bf q} \times {{\bf S}}_{-{\bf q}})  \right]
\frac{{\rm d}^3{\bf q}}{(2\pi)^3}.
\label{class_dynamics_9}
\end{eqnarray}
A very small magnetic anisotropy is neglected \cite{Hu18}. The quantity ${\bf S}_{\bf q}$ is the Fourier transform of the dimensionless manganese spin and $V$ the sample volume.

\subsection{Dispersion relation}
\label{helimagnons_dispersion}

Working in the $({\bf x}, {\bf y}, {\bf z})$ frame, the helimagnon dispersion relation is computed in the linear spin-wave approximation \cite{Maleyev06}. A gapless mode with the expected cylindrical symmetry around the direction defined by ${\bf k}$ is found:
\begin{eqnarray}
\omega({\bf q}) =  \sqrt{c_\parallel q_\parallel^2 + c_\perp {\bf q}_\perp^4},
\label{class_dynamics_22}
\end{eqnarray}
where $c_\parallel$ and $c_\perp$ are elastic constants and $q_\parallel$ and ${\bf q}_\perp$ are the ${\bf q}$ components parallel and perpendicular to ${\bf k}$ \cite{Belitz06,Ho10}. According to Ref.~\onlinecite{Maleyev06},
\begin{eqnarray}
c_\parallel  =  \left (\frac{B_1k S}{ \hbar} \right )^2  {\rm and} \,\,\,
c_\perp      =  \epsilon \frac{c_\parallel}{k^2},
\label{class_dynamics_22_1}
\end{eqnarray}
where $\epsilon$ reflects umklapp processes due to the DM interaction. The simplest approximation valid at $q=0$ and $q \gg k$, gives $\epsilon = 1/2$; see also Ref.~\onlinecite{Belitz06}. $S$ is derived from $m(T = 0) = g S\mu_{\rm B} = 0.3891\,(5) \, \mu_{\rm B}$ --- see caption of Fig.~\ref{experimental_data} --- with $g \approx 2$ \cite{Date77,Demishev11}. 

\section{Description of \lowercase{m}(T)}
\label{helimagnons_T}

The decrease of the modulus of the magnetic moment as the compound is warmed stems from the thermal variation of the helimagnon population. In the rotating reference frame, we deal with a ferromagnet albeit with the unusual dispersion relation of Eq.~\ref{class_dynamics_22}. Hence, referring e.g.\ to Ref.~\onlinecite{Kittel63}, 
\begin{eqnarray}
m(T) = m(T = 0) \left(1 - \frac{1}{N}\sum_{\bf q} n_{\bf q}  \right),
\label{moment_1}
\end{eqnarray}
where $N$ is the number of manganese atoms in the sample and $n_{\bf q}$ is the population of helimagnons with wavevector ${\bf q}$. Replacing the discrete sum by an integral,
\begin{eqnarray}
\sum_{\bf q} n_{\bf q} = V \int \frac{1}{\exp \left (\frac{\hbar \omega(\bf q)}{k_{\rm B}T} \right ) -1} 
\frac{{\rm d}^3 {\bf q}}{(2 \pi)^3}.
\label{moment_3}
\end{eqnarray}
The volume of integration which is nominally that of the first Brillouin zone is extended to the whole space since we are interested by the low temperature limit. The cylindrical symmetry of $\omega({\bf q})$ results in a $T^2$ decay of $m(T)$:
\begin{eqnarray}
  m(T) = m(T = 0) [1 - (T/T_{\rm he})^2 ],
\label{moment_4_bis}
\end{eqnarray}
with 
\begin{eqnarray}
T_{\rm he} = 4\sqrt{\frac{3}{\pi v_0 }}\frac{\hbar}{k_{\rm B}}  \left (c_\parallel c_\perp \right)^{1/4},
\label{moment_4}
\end{eqnarray}
where $v_0$ is the volume per manganese atom. From the slope of $m(T)$ versus $T^2$, the product of the parameters entering the dispersion relation can be deduced. Combining with Eq.~\ref{class_dynamics_22_1}, a relation is found between $B_1$ and $T_{\rm he}$:
\begin{equation}
  B_1 = \sqrt{\frac{\pi v_0}{3 k}} \frac{k_{\rm B} T_{\rm he}}{4\epsilon^{1/4}S}.
  \label{B1_The}
\end{equation}

\section{Description of the $\mu$SR relaxation}
\label{helimagnons_rate}

\begin{figure}
{\includegraphics[width=\linewidth]{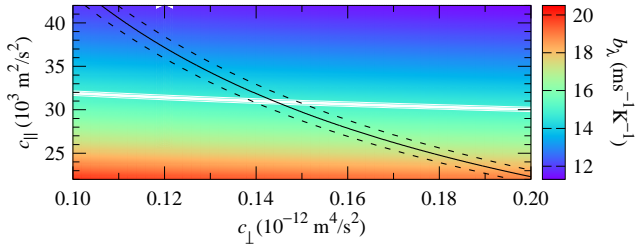}}
\caption{
Map of $b_\lambda$, resulting from Eq.~\ref{lambdaZ_dis_3_1} together with Eq.~\ref{lambdaZ_dis_3}, as a function of $c_\parallel$ and $c_\perp$. The equation for the full line is $c_\parallel c_\perp$ = $4.46 \times 10^{-9}$~m$^6$\,s$^{-4}$, and the region comprised between the dashed lines materializes the experimental uncertainty on the product $c_\parallel c_\perp$. The blank stripe across the figure shows the range for $b_\lambda$ arising from the experimental value $b_\lambda$ = 14.9\,(1)~ms$^{-1}$K$^{-1}$. Hence the blank region limited by the two dashed lines provides the experimental estimates for $c_\parallel$ and $c_\perp$.}
\label{lambda}
\end{figure}

The relaxing part of the polarization function is found, to a good approximation, to be exponential. It is characterized by a relaxation rate $\lambda_Z$ proportional to the temperature, i.e.
\begin{eqnarray}
\lambda_Z = b_\lambda T.
\label{helimagnons_rate_1}
\end{eqnarray}
The quantity $b_\lambda$ only depends on  $c_\parallel$ and $c_\perp$. The derivation of $b_\lambda$ is detailed in Appendix~\ref{Long_pol}. It is numerically computed as a function of these two parameters with the result shown in Fig.~\ref{lambda}. No other free parameter enters the computation. Remarkably, $b_\lambda$  depends weakly on $c_\perp$ but strongly on $c_\parallel$.

\section{Discussion}
\label{discussion}

The prediction of a $T^2$ decrease of $m(T)$ from its value at zero temperature (Eq.~\ref{moment_4_bis}) and the experimental observation (Fig.~\ref{experimental_data}) agree. This result provides a welcome support for the theoretical expression of the dispersion relation (Eq.~\ref{class_dynamics_22}). Similarly, Eq.~\ref{helimagnons_rate_1} offers an explanation of the experimental observation of $\lambda_Z(T) \propto T$ \cite{Yaouanc05}.

Quantitatively, from $T_{\rm he} = T_{\rm c}/\sqrt{a} = 50.2\,(5)$~K --- see caption of Fig.~\ref{experimental_data} --- we compute $c_\parallel c_\perp = 4.46\,(15) \times 10^{-9}~{\rm m}^6\,{\rm s}^{-4}$. Combining the experimental determination of $b_\lambda$ with that of $T_{\rm he}$ and taking advantage of the different dependencies of these quantities on the dispersion relation parameters, we can independently extract $c_\parallel$ and $c_\perp$ which are found temperature independent up to $\approx 18$~K. From Fig.~\ref{lambda} we derive $c_\perp = 0.144\,(8) \times 10^{-12}$~m$^4$\,s$^{-2}$ and $c_\parallel = 31.0\,(7) \times 10^3$~m$^2$\,s$^{-2}$.

Neutron scattering measurements give $c_\perp = 0.11\,(1) \times 10^{-12}$~m$^4$/s$^2$ at 25-26~K \cite{Semadeni99,Sato16}. Hence, as expected, approaching $T_{\rm c}$ from below a weak softening of $c_\perp$ occurs. Moreover, these data were recorded at relatively large momentum transfer \cite{Kugler15}. Our $\lambda_Z(T)$ data probe excitations at much lower energy (see, e.g.\ Fig.~\ref{correlation} below), indicating the validity of Eq.~\ref{class_dynamics_22} over a broad energy range. From our data, we deduce $\epsilon = c_\perp k^2/c_\parallel = 0.57\,(5)$. This is in line with the estimate of Refs.~\onlinecite{Maleyev06,Belitz06}  --- see Eq.~\ref{class_dynamics_22_1} and the discussion below it. From the $T_{\rm he}$ value, using Eq.~\ref{B1_The} we compute $B_1  = 2.73\, (4) \times 10^{-40}$~J\,m$^2$ and therefore $|D| =  k B_1 = 9.54 \, (10) \times 10^{-32}$ J\,m. 

We have found that the combined quantitative analysis of $m(T)$ and $\lambda_Z(T)$  at low temperature can be performed within the framework offered by the helimagnon excitations which derive from a localised electron picture. The two parameters entering the expression of $\omega({\bf q})$ are measured. The SCR theory combined analysis fails. However, this theory successfully reproduced other physical parameters; see Sec.~\ref{Intro}. This state of affairs suggests MnSi to be a dual system where an electronic subset of itinerant states coexists with the subset of localized magnetic $3d$ electrons whose signature are the spin waves. Such a picture was introduced for the superconductor ferromagnet UGe$_2$ based on experimental data \cite{Yaouanc02,Sakarya10} --- see also \footnote{A recent work using a new neutron technique \cite{Haslbeck19} confirms the $\mu$SR results.} --- and subsequently supported by theoretical works \cite{Mineev13,Chubukov14}. We suggest MnSi to be the first experimentally recognized $3d$ dual system.

We note that a spin gap of 1.29~meV has been extracted from a specific heat measurement \cite{Mishra16}. If such a gap were present we would have detected it. The specific heat analysis requires to take into account three contributions in a relatively large temperature range: conduction electrons, phonons, and helimagnons. Hence, the information extracted from the analysis strongly depends on their modelling. On the other hand, only helimagnons matter for $m(T)$ and $\lambda_Z(T)$.

The methodology exposed in this report can be applied to other helimagnets in particular when the low-temperature zero-field magnetic moment decays as $T^2$. The insulating magnetoelectric system Cu$_2$OSeO$_3$ \cite{Bos08} is such a system \cite{Maisuradze11,Adams12} and will be discussed in a separate paper.

\section{Conclusions and comments}
\label{conclusion}

We have shown the helimagnon dispersion to quantitatively explain $\lambda_Z(T)$ and $m(T)$ which probe the sub-$10^{-7}$~eV and $10^{-3}$~eV energy range, respectively. Hence, this dispersion is valid over a huge energy range. There is no spin-gap at the sub-$10^{-7}$~eV level since otherwise energy conservation during the relaxation process would not be fulfilled. The SCR theory does not predict the observed $m(T)$, suggesting the compound to be a dual system. 

Remarkably, $m(T)$ and $\lambda_Z (T)$ are described in a continuum framework, while we already know that such an approach cannot explain the MnSi magnetic structure \cite{Dalmas16}. Solving this apparent contradiction is left for a future study.

Obviously, the methodology used in this report for the characterization of magnetic excitations in zero field at small ${\bf q}$ can be applied to other helimagnets.

After the submission of this work we became aware of two relevant  works. Choi {\it et al.} \cite{Choi19} proposed MnSi to be a dual system based on theoretical arguments. Consistent with this proposal, the analysis of neutron scattering data with density functional theory (DFT) paired with dynamical mean-field theory (DMFT) suggests MnSi to display strong local electron correlations driven by Hund's coupling \cite{Chen19}. Our data support the dual picture of these two works. 

\begin{acknowledgments}
We thank M. Janoschek for a careful reading of the manuscript and insightful comments and I. Mirebeau for useful discussions. Part of this work was performed at the GPS spectrometer of the Swiss Muon Source (Paul Scherrer Institute, Villigen, Switzerland). D.A.\ acknowledges partial financial support from the Romanian UEFISCDI project PN-III-P4-ID-PCE-2016-0534. 
\end{acknowledgments}

\appendix

\section{Longitudinal $\mu$SR polarization function}
\label{Long_pol}

Here we describe the relaxing $\mu$SR polarization function in terms of helimagnon excitations. This function is relevant for the description of the muon spin-lattice relaxation. We first establish the expressions of the spin-correlation functions needed for the computation of the spin-lattice relaxation rate. Then we expose a detailed treatment of this rate. This enables us to compute the relaxing part of the measured polarization function and show that the measured relaxation rate is proportional to the temperature. 

\subsection{Spin correlation functions}
\label{Long_pol_corr}

The derivation of an expression for the spin relaxation rate $\lambda_Z$ (see Sec.~\ref{Long_pol_rate}) requires the symmetrized spin correlation tensor, a component of which is defined as
\begin{eqnarray}
\Lambda^{\alpha\beta}({\bf q},\omega) & = & \frac{1}{2} \left [\langle S^\alpha({\bf q},\omega) S^\beta(-{\bf q}) \rangle  
+ \langle S^\beta(-{\bf q}) S^\alpha({\bf q},\omega)\rangle\right ], \cr  &
\label{Quantum_corr_1}
\end{eqnarray}
where $\alpha$ and $\beta$ are Cartesian axis labels. As seen in Sec.~\ref{Long_pol_rate}, correlations in the laboratory reference frame are needed since the measurement is obviously performed in this frame. However their derivation from the Hamiltonian is more conveniently obtained in the rotating reference frame associated with each magnetic domain. The result is given in Sec.~\ref{Long_pol_corr_rotating} where we immediately identify contrasting ${\bf q}$ dependencies for the different tensor components. In the subsequent section we provide the expression of the spin correlations in the magnetic domain frame.

\subsubsection{Correlations in the rotating frame}
\label{Long_pol_corr_rotating}

The rotating reference frame is defined in Eq.~\ref{rotating} in terms of the $({\bf a}, {\bf b}, {\bf n})$ frame. As usual in spin wave theory it is convenient to introduce the spin raising and lowering operators $S^\pm = S^x \pm i S^y$. Referring to the $(+,-,z)$ basis we need {\sl a priori} to consider nine symmetrized spin correlation functions: $\Lambda^{+z}({\bf q},\omega)$, $\Lambda^{-z}({\bf q},\omega)$, $\Lambda^{z-}({\bf q},\omega)$, $\Lambda^{z+}({\bf q},\omega)$, $\Lambda^{zz}({\bf q},\omega)$, $\Lambda^{+-}({\bf q},\omega)$, $\Lambda^{++}({\bf q},\omega)$, $\Lambda^{-+}({\bf q},\omega)$, and $\Lambda^{--}({\bf q},\omega)$. The first four functions vanish because they involve the product of three helimagnon operators, the thermal average of which is zero. The longitudinal correlation $\Lambda^{zz}_{\bf q}(\omega)$ accounts for the Raman scattering process, i.e.\ a two magnon process \cite{Dalmas95}. It is negligible here relative to the single magnon scattering process \footnote{The muon spin relaxation rate associated to magnon Raman scattering is estimated to be $10^{-5}$~$\mu$s$^{-1}$ at 10~K; see Eq. 10.218 in Ref.~\onlinecite{Yaouanc11}. The spin wave stiffness constant $D_{\rm FM}$ therein is directly related to $c_\perp$.} which plays a crucial role due to the gapless nature of the helimagnon dispersion. All this is well known from old nuclear magnetic resonance works \cite{Moriya56a,Moriya56b,vanKranendonk56,Mitchell57,Beeman68}. 

We are left with $\Lambda^{+-}({\bf q},\omega)$, $\Lambda^{++}({\bf q},\omega)$, $\Lambda^{-+}({\bf q},\omega)$, and $\Lambda^{--}({\bf q},\omega)$. These quantities are readily computed after the diagonalization of the Hamiltonian (Eq.~\ref{Quantum_dis_3}). We find it convenient to express the correlations in the Cartesian coordinates $\{x,y,z\}$, with the result
\begin{eqnarray}
\Lambda^{xx}({\bf q},\omega)  & = & \frac{\pi k_{\rm B} T}{B_1(q^2+k^2)} 
\left\{ \delta[\omega-\omega({\bf q})]+\delta[\omega+\omega({\bf q})]\right\} , \cr
\Lambda^{yy}({\bf q},\omega)  & = & \frac{\pi  k_{\rm B} T}{B_1 q^2} \left\{\delta[\omega-\omega({\bf q})]+ \delta(\omega+\omega({\bf q})]\right\}, \label{Quantum_corr_8} \\
\Lambda^{xy}({\bf q},\omega) & = & 
\frac{\pi  k_{\rm B} T}{i B_1 q\sqrt{q^2+k^2}} \left\{\delta[\omega-\omega({\bf q})]- \delta[\omega+\omega({\bf q})]\right\},
\nonumber
\end{eqnarray}
and $\Lambda^{yx}({\bf q},\omega)  =  -\Lambda^{xy}({\bf q},\omega)$. It occurs that only the first two correlations will be of use (Eq.~\ref{lambdaZ_single_7}). This particularly simple form for the correlations has been obtained using the isotropic dispersion (Eq.~\ref{class_dynamics_20}). Nevertheless, the expression of $\omega({\bf q})$ reflecting the actual cylindrical symmetry around ${\bf k}$ (Eq.~\ref{class_dynamics_22}) will be kept in the arguments of the Dirac delta functions. This is in line with the fact that in first-order perturbation theory, a perturbation, such as the DM interaction, influences the eigenvalues but not the eigenstates of an operator. This is essential for the computation of $\lambda_Z(T)$. 

The correlations $\Lambda^{xx}({\bf q},\omega)$ and $\Lambda^{yy}({\bf q},\omega)$ differ by their $q$-dependence:  the latter diverges when ${\bf q}$ approaches 0 while the former does not. This suggests that the contribution of $\Lambda^{yy}({\bf q},\omega)$ to $\lambda_Z$ will be dominant.  

\begin{figure}
{\includegraphics[width=\linewidth]{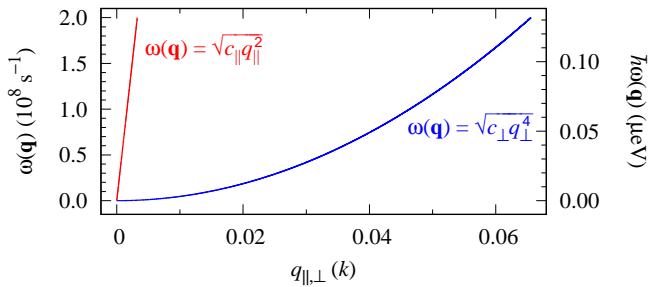}}
\caption{
Dispersion relation for ${\bf q}$ along the direction defined by ${\bf k}$ and the direction perpendicular to it, restricted to the energy range of interest for a muon spin relaxation process involving the creation or annihilation of a helimagnon. The curves are calculated for the $c_\parallel$ and $c_\perp$ values found in this study. Notice the strong anisotropy and the extremely small wavevector scale which is expressed in unit of $k$. Angular frequency and energy scales are provided.}
\label{correlation}
\end{figure}

Let us now examine the order of magnitude of the relevant $\omega$ values and its implication. From the highest value of the field at the muon site (Fig.~\ref{Fig_distributions}), we have $|\omega| < 2\times 10^8$~s$^{-1}$. Therefore, $\omega({\bf q})$ must be very small to obey the conservation of energy enforced by the Dirac delta functions in Eq.~\ref{Quantum_corr_8}: only modes at extremely small energy matter. This justifies the limit $k_{\rm B} T \gg \hbar \omega({\bf q})$ taken for the prefactors derivation in Eq.~\ref{Quantum_corr_8}. For definiteness, the range of wavevectors explored by the correlation functions is illustrated in Fig.~\ref{correlation}.

\subsubsection{Correlations in the magnetic domain frame}
\label{Long_pol_corr_magnetic}

The correlations can now be expressed in the $({\bf a}, {\bf b}, {\bf n})$ frame. This is achieved using the $R_{{\bf k}_\ell\cdot {\bf r}}$ rotation matrix of Appendix.~\ref{frames}. The finite elements of the correlation tensor are
\begin{eqnarray}
\Lambda^{a a } ({\bf q}, \omega) &  = & 
\frac{1}{4} \left [\Lambda^{xx} ({\bf q} - {\bf k},\omega) + \Lambda^{xx} ({\bf q} + {\bf k},\omega)  \right ]\cr
\Lambda^{b b } ({\bf q},\omega) &  = & \Lambda^{a a } ({\bf q}, \omega)
,\cr
\Lambda^{a b } ({\bf q}, \omega) &  = & 
\frac{1}{4i} \left [\Lambda^{xx} ({\bf q} - {\bf k},\omega) - \Lambda^{xx} ({\bf q} + {\bf k},\omega)  \right ].\cr
\Lambda^{b a } ({\bf q},\omega) &  = & -\Lambda^{a b } ({\bf q}, \omega),\cr
\Lambda^{nn} ({\bf q},\omega) & \equiv & \Lambda^{yy}({\bf q},\omega), \label{lambdaZ_single_7} 
\end{eqnarray}
The final line arises from the fact that the rotation is about ${\bf n} \equiv {\bf y}$.

Equation \ref{lambdaZ_single_7} together with Eq.~\ref{Quantum_corr_8} and Fig.~\ref{correlation} imply that only regions in the very immediate vicinity of ${\bf q} = 0$ and ${\bf q}$ = $\pm {\bf k}$ contribute to $\Lambda^{n n }({\bf q},\omega)$ and the other correlation functions, respectively \cite{Yosida61,Cooper62}. The latter correlations correspond to phasons \cite{Izyumov85}, in agreement with their planar nature (normal to ${\bf k}$) \cite{Overhauser71}.

\subsection{Muon spin-lattice relaxation rate}
\label{Long_pol_rate}

We proceed with the expression of the relaxation rate $\lambda_Z$. We need to deal with peculiar features of MnSi. (i) To a crystallographic $4a$ muon position \cite{Amato14} correspond four magnetic sites labelled with the subscript $s$; (ii) a field distribution associated with the magnetic incommensurability is observed for three of them. In addition, four K-domains defined by symmetry equivalent ${\bf k}_\ell$ propagation wavevectors exist. Hence, the relaxation rate associated with a muon at site $s$ in domain $\ell$ is $\lambda_{Z;s,\ell} (\omega)$ where $\omega = \gamma_\mu {\bf B}_{0,\ell,s}\cdot {\bf Z}$, with $\gamma_\mu = 8.51616 \times 10^8 \, {\rm rad \, s}^{-1} {\rm T}^{-1}$ is the muon gyromagnetic ratio and ${\bf B}_{0,\ell,s}$ is the spontaneous field. We have \cite{Yaouanc11}:
\begin{eqnarray}
\lambda_{Z;s,\ell} (\omega) = \frac{{\mathcal D}}{2}
\int \sum_{\gamma,\gamma^\prime} {\mathcal A}_s^{\gamma \gamma^\prime}({\bf q})\Lambda_\ell^{\gamma \gamma^\prime}({\bf q},\omega) \frac{{\rm d}^3 {\bf q}}{(2 \pi)^3},
\label{lambdaZ_single_1}
\end{eqnarray}
with
\begin{eqnarray}
{\mathcal A}_s^{\gamma \gamma^\prime}({\bf q})  = 
\sum_{d,d^\prime} G_{d,s}^{X \gamma}({\bf q}) G_{d^\prime,s}^{X \gamma^\prime}(-{\bf q}) +
G_{d,s}^{Y \gamma}({\bf q}) G_{d^\prime,s}^{Y \gamma^\prime}(-{\bf q}).\cr & &
\label{lambdaZ_single_2}
\end{eqnarray}
Here, $\gamma,\gamma^\prime \in \{X,Y,Z\}$, the constant ${\mathcal D} = \left ({\mu_0}/{4 \pi}  \right )^2 \gamma_\mu^2 {(g\mu_{\rm B})^2}/{v_{\rm c}}$, and $v_{\rm c}$ is the unit cell volume. Tensor $\boldsymbol{\it G}_{d,s}({\bf q})$ describes the coupling between the muon spin at site $s$ and the manganese spin at crystal position $d$ in the unit cell. While this coupling is fully accounted for, $\boldsymbol{\it \Lambda}_\ell({\bf q},\omega)$ is assumed to be independent of $d$, in line with our description of the correlations \cite{Bak80,Nakanishi80}.

Through their dependence on $\boldsymbol{\it G}_{d,s}({\bf q})$, the tensor elements ${\mathcal A}_s^{\gamma \gamma^\prime}({\bf q})$ in Eq.~\ref{lambdaZ_single_1} are readily expressed in the $({\bf X}, {\bf Y}, {\bf Z})$ frame. We need explicit $\Lambda_\ell^{\gamma \gamma^\prime}({\bf q},\omega)$ in this same frame. This is achieved through the relation 
\begin{eqnarray}
\Lambda^{\gamma \gamma^\prime}_\ell({\bf q},\omega) 
& = & \sum_{\varepsilon\varepsilon^\prime}  R_\ell^{\gamma\varepsilon} \Lambda^{\varepsilon \varepsilon^\prime}({\bf q},\omega) R_\ell^{\gamma^\prime\varepsilon^\prime},
\label{correlation_domains_1}
\end{eqnarray}
where $R_\ell$ are rotation matrices --- see Appendix~\ref{matrices_R} --- and $\varepsilon$ and $\varepsilon^\prime$ label Cartesian components in the $({\bf a}, {\bf b}, {\bf n})$ magnetic domain frame in which $\Lambda^{\varepsilon \varepsilon^\prime}({\bf q},\omega)$ is naturally independent of $\ell$, while it is not in the laboratory frame.

For the evaluation of Eq.~\ref{lambdaZ_single_1} simplifications apply. Numerically, it is found that the terms involving the $\Lambda^{n n} ({\bf q}, \omega)$ spin correlations are larger by a factor at least 500 than the terms with the other correlations, see Appendix~\ref{Correlation_phasons}. This implies that the measured relaxation is dominated by modes close to ${\bf q} = {\bf 0}$ rather than close to ${\bf q}$ = $\pm{\bf k}_\ell$.
Retaining only the $n n $ correlation we get:
\begin{eqnarray}
\Lambda^{\varepsilon \varepsilon^\prime}({\bf q},\omega) & = & \Lambda^{nn}({\bf q},\omega) 
\delta_{\varepsilon,n}\delta_{\varepsilon^\prime,n} =
\Lambda^{nn}({\bf q},\omega) M_n^{\varepsilon \varepsilon^\prime}, \cr
& & {\rm with }\hspace{1mm}
M_n  =  \begin{pmatrix} 0 & 0 & 0 \cr 0 & 0 & 0 \cr 0 & 0 & 1 \end{pmatrix}.
\label{correlation_domains_4}
\end{eqnarray}
With this result, Eq.~\ref{correlation_domains_1} drastically simplifies to
\begin{eqnarray}
\Lambda^{\gamma \gamma^\prime}_\ell({\bf q},\omega) 
 =  \Lambda^{nn}({\bf q},\omega) N_{n,\ell}^{\gamma \gamma^\prime}; \, N_{n, \ell} = R_\ell M_n R_\ell^{-1}.
\label{correlation_domains_1_2}
\end{eqnarray}

As the spin relaxation is driven by modes in the vicinity of ${\bf q} = 0$, we are entitled to consider the long-wavelength limit of $G_{d,s}^{\alpha \beta}({\bf q})$. Referring to Refs.~\onlinecite{Yaouanc93,Yaouanc93a,Dalmas94,Frey97,Yaouanc96},
\begin{eqnarray} 
G_{d,s}^{\alpha \beta}({\bf q}) =- 4 \pi \left (P^{\alpha \beta}_{\rm L} ({\bf q}) -C^{\alpha \beta}_{d,s} ({\bf 0})
-\frac{r_\mu H}{4 \pi} \delta^{\alpha \beta}  
\right ),
\label{lambdaZ_single_13}
\end{eqnarray}
where $P^{\alpha \beta}_{\rm L}({\bf q})  = q^\alpha q^\beta/q^2 $ is an element of the longitudinal projection operator and $r_\mu H/4 \pi$ a hyperfine coupling constant. The tensor $\boldsymbol{\it C}_{d,s}({\bf 0})$ describes the analytical part at ${\bf q} = 0$ of the dipole interaction between the muon and manganese spins. 

Because $P^{\alpha \beta}_{\rm L}({\bf q})$ depends on the polar and azimuthal angles, $\theta$ and $\varphi$ respectively, of ${\bf q}$ in the $({\bf X},{\bf Y},{\bf Z})$ frame, but not on $q$, the evaluation of $\lambda_{Z;s,\ell} (\omega)$ is conveniently performed in spherical coordinates. Since conversely $\Lambda^{nn}({\bf q},\omega)$ depends only on $q$ and $\theta$, the $\varphi$ integration in Eq.~\ref{lambdaZ_single_1} is restricted to the coupling tensor elements. Hence, we find it convenient to introduce the geometrical functions ${\mathcal V}^{\gamma\gamma^\prime}_s (\theta)$ such that
\begin{eqnarray}
\int_0^{2\pi} {\mathcal A}^{\gamma\gamma^\prime}_s({\bf q}) {\rm d}\varphi & = & 32 \pi^3 {\mathcal V}^{\gamma\gamma^\prime}_s (\theta).
\label{Int_gamma_gamma_prime}
\end{eqnarray}
The angular functions ${\mathcal V}^{\gamma \gamma^\prime}_s(\theta)$ are listed in Appendix~\ref{functions}. Thanks to the presence of Dirac delta functions in the expression of $\Lambda^{nn}({\bf q},\omega)$, the integration over $q$ is analytical. Finally, we derive
\begin{eqnarray}
\lambda_{Z;s,\ell} (\omega)  =  \alpha_{s,\ell} (\omega) T,
\label{lambdaZ_single_1_bis}
\end{eqnarray}
where
\begin{eqnarray}
\alpha_{s,\ell} (\omega)  =  \frac{ \pi {\mathcal D} k_{\rm B} k S}{2 \hbar c^{1/2}_\parallel}  \sum_{\gamma,\gamma^\prime}{\mathcal C}^{\gamma\gamma^\prime}_s(\omega) N_{n,\ell}^{\gamma \gamma^\prime}.
\label{lambdaZ_single_1_three_three}
\end{eqnarray}
We have introduced the functions
\begin{eqnarray}
{\mathcal C}^{\gamma \gamma^\prime}_s(\omega) = 8
\int_0^\pi \frac{\sqrt{c_\parallel\cos^2 \theta + c_\perp \sin^4 \theta q^2_0}}
{c_\parallel\cos^2 \theta + 2  c_\perp \sin^4 \theta q^2_0}
{\mathcal V}^{\gamma \gamma^\prime}_s(\theta) \sin \theta{\rm d} \theta , \cr & &
\label{Int_3_bis}
\end{eqnarray}
and defined the auxiliary function $q^2_0(\theta,\omega)$: 
\begin{eqnarray}
q^2_0(\theta,\omega) = \frac{}{} \frac{-c_\parallel\cos^2 \theta + \sqrt{ c_\parallel^2\cos^4 \theta + 4 c_\perp \omega^2 \sin^4 \theta}}
{2c_\perp \sin^4 \theta}.\cr & &
\label{integral_6}
\end{eqnarray}
The polar angle integration in Eq.~\ref{Int_3_bis} is numerical. From Eq.~\ref{lambdaZ_single_1_bis} we find $\lambda_{Z;s,\ell}(\omega)$ to be proportional to $T$.

\subsection{Longitudinal polarization function}
\label{Long_pol_func}

Denoting as ${\bf r}_{0,s}$ the vector distance between site $s$ in a given cubic cell and the origin of the crystal and $n_{\rm c}$ the number of unit cells, following Eq.~11 of Ref.~\onlinecite{Dalmas16} the spontaneous field at this muon site is $\boldsymbol{\mathfrak{B}}_{0,\ell,{\bf k}_\ell,s}(-{\bf k}_\ell\cdot {\bf r}_{0,s}) +$ c.c.\ with the definition 
\begin{eqnarray}
\boldsymbol{\mathfrak{B}}_{0,\ell,{\bf q},{\rm s}}(\psi) & = & \frac{\mu_0}{4\pi} \frac{g\mu_{\rm B}}{\sqrt{n_{\rm c}}v_{\rm c}}\sum_d {\boldsymbol G}_{d,{\rm s}}({\bf q}){\bf S}_{\ell,d}({\bf q})\exp(i\psi). \cr
& & 
\label{lambdaZ_dis_1}
\end{eqnarray}
Because in a $\mu$SR experiment several millions of muons are implanted randomly in different cells of the crystal, it is a very good approximation to state that the muons at site $s$ in domain $\ell$  probe the continuous field distribution
\begin{eqnarray}
{\bf B}_{0,\ell,{\rm s}}(\psi) & = & \boldsymbol{\mathfrak{B}}_{0,\ell,{\bf k}_\ell,{\rm s}}(\psi) + {\rm c.c.}
\label{lambdaZ_dis_2}
\end{eqnarray}
where the variable $\psi$ spans the interval $[0, 2\pi[$.

The relaxing part of the polarization function is \cite{Yaouanc11}
\begin{eqnarray}
P_Z^{\rm rel}(t,T) & = & \frac{1}{16} \sum_{\ell, s} \int_0^{2\pi} 
\left({\bf {\hat B}}_{0,\ell,s}(\psi)\cdot {\bf Z}\right)^2 
\label{lambdaZ_dis_3} \\
& & \hspace{10mm} \times \exp \left [ -\alpha_{s,\ell}(\gamma_\mu {\bf B}_{0,\ell,s}(\psi)\cdot {\bf Z} ) T t \right ] {\rm d} \psi,
\nonumber
\end{eqnarray}
where ${\bf {\hat B}}_{0,\ell,s}(\psi)$ is the unit vector parallel to ${\bf B}_{0,\ell,s}(\psi)$. The normalization factor 16 is the product of the numbers of magnetic domains and muon sites in the unit cell. To a good approximation, $P_Z^{\rm rel}(t,T)$ is found numerically to be an exponential function of time:
\begin{eqnarray}
P_Z^{\rm rel}(t,T)  =  P_Z^{\rm rel}(Tt) =\frac{1}{3} \exp \left ( -\lambda_Z t \right );  \, \lambda_Z = b_\lambda T. \cr
& & 
\label{lambdaZ_dis_3_1}
\end{eqnarray}
The quantity $b_\lambda$ is numerically computed as a function of $c_\parallel$ and $c_\perp$ as explained in Appendix~\ref{polarization}. The result is presented in the main text; see Sec.~\ref{helimagnons_rate}.

\section{Additional information on helimagnon excitations}
\label{Aux}

Calculations presented in Appendix~\ref{Long_pol} require miscellaneous results which, for the sake of completeness and simplicity, are grouped in this appendix.

Equation \ref{lambdaZ_single_1} provides the formula for the relaxation rate. It requires an expression for the spin correlation functions in the laboratory reference frame $({\bf X}, {\bf Y}, {\bf Z})$. To get this expression, we start writing down the Hamiltionian, and express the correlation functions in the rotating frame $({\bf x}, {\bf y}, {\bf z})$. Then we provide the rotation matrix for a transformation from the rotating frame to the magnetic domain frame $({\bf a}, {\bf b}, {\bf n})$. We recall that the latter frame is attached to the magnetic domain (${\bf n} \equiv {\bf k}_\ell/k_\ell$) while the former is the frame rotating with the helical structure. In the subsequent subsection rotation matrices which account for the magnetic domains are given. In the fourth subsection the phason contribution to the relaxation is spelt out. Angular functions  written in the $({\bf X}, {\bf Y}, {\bf Z})$ frame and introduced to describe the relaxation induced by the fluctuations in the $q = 0$ vicinity are then listed. Information on the numerical computation of the relaxing part of the polarization function follows. In the last subsection final comments are given.

\subsection{Hamiltonian}
\label{Hamiltonian}

In line with the helical structure of MnSi we consider a reference frame $({\bf x}({\bf r}), {\bf y}({\bf r}), {\bf z}({\bf r}))$, rotating with the Mn moment direction:
\begin{eqnarray}
  {\bf z}({\bf r}) & = & \cos({\bf k} \cdot {\bf r})\, {\bf a} - \sin({\bf k} \cdot {\bf r})\, {\bf b}, \cr
  {\bf x}({\bf r}) & = & \sin({\bf k} \cdot {\bf r})\, {\bf a} + \cos({\bf k} \cdot {\bf r})\, {\bf b}, \cr
  {\bf y}({\bf r}) & = & {\bf k}/k \equiv {\bf n}.
\label{rotating}
\end{eqnarray}
Since we are in a continuous description, we have dropped the $d$ subscripts for ${\bf a}$ and ${\bf b}$. In the following, for the ease of notation we will drop the ${\bf r}$ argument to ${\bf x}$, ${\bf y}$ and ${\bf z}$.
Starting from the energy expression given in Eq.~\ref{class_dynamics_9} and introducing the aforementioned spin raising and lowering operators and the boson operators through the relations $S_{\bf q}^+ = \sqrt{2S} a^\dag_{-{\bf q}}$ and $S_{\bf q}^- = \sqrt{2S} a_{\bf q}$, we write down the Heisenberg Hamiltonian ${\mathcal E}_{\rm H}$. Keeping only terms bilinear in boson operators, we derive
\begin{eqnarray}
{\mathcal E}_{\rm H} & = & \sum_{\bf q} 
 \left[L_q a^\dag_{\bf q}a_{\bf q} + \frac{W_{\bf q}}{2} \left (a_{\bf q}a_{-{\bf q}} + a^\dag_{\bf q}a^\dag_{-{\bf q}} \right )\right],
\label{Quantum_dis_3}
\end{eqnarray}
with $L_q = \frac{B_1S}{2}(2 q^2 + k^2)$ and $W_{\bf q} = W  = \frac{B_1S}{2} k^2$. The expression for ${\mathcal E}_{\rm H}$ neglects the Dzyaloshinski-Moriya (DM) exchange interactions. Its presence is nevertheless acknowledged by the underlying rotating frame.

The Hamiltonian of Eq.~\ref{Quantum_dis_3} is diagonalized using a conventional Bogoliubov transformation. The excitation dispersion relation is
\begin{eqnarray}
\hbar\omega({\bf q}) =  B_1S  \sqrt{k^2q^2 + q^4}.
\label{class_dynamics_20}
\end{eqnarray}

\subsection{From the $({\bf x}, {\bf y}, {\bf z})$ to the $({\bf a}, {\bf b}, {\bf n})$ frames}
\label{frames}

While in Eq.~\ref{Quantum_corr_8} the tensor elements of $\boldsymbol{\it \Lambda}_\ell({\bf q},\omega)$ are written in the $({\bf x}, {\bf y}, {\bf z})$ frame, their counterparts in the $({\bf a}, {\bf b}, {\bf n})$ frame are needed. Introducing the rotation matrix
\begin{eqnarray}
R_{{\bf k}_\ell\cdot {\bf r}} = 
\begin{pmatrix}
    \sin({\bf k}_\ell \cdot {\bf r}) & 0 & \cos({\bf k}_\ell \cdot {\bf r}) \cr
    \cos({\bf k}_\ell \cdot {\bf r}) & 0 & -\sin({\bf k}_\ell \cdot {\bf r}) \cr
     0                           &  1                          &  0 \cr
\end{pmatrix},
\label{lambdaZ_single_3}
\end{eqnarray}
in the direct space we can express a spin component in the $({\bf a}, {\bf b}, {\bf n})$ frame in terms of spin components in the $({\bf x}, {\bf y}, {\bf z})$ frame such as 
\begin{eqnarray}
S^\epsilon ({\bf r}) = \sum_\rho R^{\epsilon \rho} _{{\bf k}_\ell\cdot {\bf r}} S^\rho ({\bf r}).
\label{lambdaZ_single_4}
\end{eqnarray}
Therefore 
\begin{eqnarray}
\Lambda^{\epsilon \epsilon^\prime}_\ell({\bf r}, {\bf r}^\prime;\omega) = 
\sum_{\rho, \rho^\prime} R^{\epsilon \rho} _{{\bf k}_\ell\cdot {\bf r}} R^{\epsilon^\prime \rho^\prime} _{{\bf k}_\ell\cdot {\bf r}^\prime }
\Lambda_\ell^{\rho \rho^\prime}({\bf r}, {\bf r}^\prime;\omega),
\label{lambdaZ_single_5}
\end{eqnarray}
with $\epsilon,\epsilon^\prime \in \{a,b,n\}$. Since
\begin{eqnarray}
\Lambda^{\epsilon \epsilon^\prime}_\ell({\bf q},\omega) = \frac{1}{n_{\rm c}}
\sum_{{\bf r}, {\bf r}^\prime} \exp \left [-i {\bf q} \cdot \left ({\bf r} - {\bf r}^\prime  \right )  \right ]
\Lambda^{\epsilon\epsilon^\prime}({\bf r}, {\bf r}^\prime;\omega), \cr
& & 
\label{lambdaZ_single_6}
\end{eqnarray}
the correlation functions in the $({\bf a}, {\bf b}, {\bf n})$ can be computed.

\subsection{Rotation matrices $R_\ell$}
\label{matrices_R}

Here we discuss the rotation matrices $R_\ell$ which are introduced in Eq.~\ref{correlation_domains_1}.

The $R_\ell$ matrices are conveniently written as products of $R(\varphi,\theta,\psi)$ matrices which describe the rotation from crystal coordinates to magnetic domains coordinates \footnote{The notation for $R(\varphi,\theta,\psi)$ follows that of Eq.~D6 in Ref.~\onlinecite{Yaouanc11}}, namely $R_{{\rm c}\to 111}\equiv R(\pi/4,\theta_+,0)$, $R_{{\rm c}\to \bar{1}\bar{1}1} \equiv R(5\pi/4,\theta_+,0)$, $R_{{\rm c}\to \bar{1}1\bar{1}} \equiv R(3\pi/4,\theta_-,0)$, and $R_{{\rm c}\to 1\bar{1}\bar{1}} \equiv R(7\pi/4,\theta_-,0)$, with $\theta_\pm = \arccos{(\pm 1/\sqrt{3})}$. This means that
\begin{eqnarray}
  R_{{\rm c}\to 111} & = &
  \begin{pmatrix}
    \frac{1}{\sqrt{6}}  & \frac{1}{\sqrt{6}} & \frac{-2}{\sqrt{6}} \\[1.5mm]
    \frac{-1}{\sqrt{2}} & \frac{1}{\sqrt{2}} & 0 \\[1.5mm]
    \frac{1}{\sqrt{3}}  & \frac{1}{\sqrt{3}} & \frac{1}{\sqrt{3}}
  \end{pmatrix}, 
  R_{{\rm c}\to \bar{1}\bar{1}1} =
  \begin{pmatrix}
    \frac{-1}{\sqrt{6}} & \frac{-1}{\sqrt{6}} & \frac{-2}{\sqrt{6}} \\[1.5mm]
    \frac{1}{\sqrt{2}}  & \frac{-1}{\sqrt{2}} & 0 \\[1.5mm]
    \frac{-1}{\sqrt{3}} & \frac{-1}{\sqrt{3}} & \frac{1}{\sqrt{3}}
  \end{pmatrix},\nonumber \\[2mm]
  R_{{\rm c}\to \bar{1}1\bar{1}} & = &
  \begin{pmatrix}
    \frac{1}{\sqrt{6}}  & \frac{-1}{\sqrt{6}} & \frac{-2}{\sqrt{6}} \\[1.5mm]
    \frac{-1}{\sqrt{2}} & \frac{-1}{\sqrt{2}} & 0 \\[1.5mm]
    \frac{-1}{\sqrt{3}}  & \frac{1}{\sqrt{3}} & \frac{-1}{\sqrt{3}}
  \end{pmatrix}, 
  R_{{\rm c}\to 1\bar{1}\bar{1}} =
  \begin{pmatrix}
    \frac{-1}{\sqrt{6}} & \frac{1}{\sqrt{6}} & \frac{-2}{\sqrt{6}} \\[1.5mm]
    \frac{1}{\sqrt{2}}  & \frac{1}{\sqrt{2}} & 0 \\[1.5mm]
    \frac{1}{\sqrt{3}} & \frac{-1}{\sqrt{3}} & \frac{-1}{\sqrt{3}}
  \end{pmatrix}.\cr & &
\label{correlation_domains_2}
\end{eqnarray}
With these definitions,
$R_{\ell = \bar{1}\bar{1}1}$ = $R_{{\rm c}\to 111} R_{{\rm c}\to \bar{1}\bar{1}1}^{-1}$,
$R_{\ell = \bar{1}1\bar{1}}$ = $R_{{\rm c}\to 111} R_{{\rm c}\to \bar{1}1\bar{1}}^{-1}$,
$R_{\ell = 1\bar{1}\bar{1}}$ = $R_{{\rm c}\to 111} R_{{\rm c}\to 1\bar{1}\bar{1}}^{-1}$, and obviously $R_{\ell = 111}$ is equal to the identity.

\subsection{Contribution of the phason correlations to the relaxation}
\label{Correlation_phasons}

Here we discuss the contribution to the relaxation of the spin correlations normal to the propagation wavevector. 

From the relations resulting from Eq.~\ref{lambdaZ_single_7}, the correlation tensor expressed in the magnetic domain frame can be summarized in the synthetic form
\begin{eqnarray}
  \Lambda^{aa}({\bf q},\omega) M_a + \Lambda^{ab}({\bf q},\omega) M_b +
  \Lambda^{nn}({\bf q},\omega) M_n,
 \label{correlation_domains_4_bis}
\end{eqnarray}
where
\begin{eqnarray}
  M_a = \begin{pmatrix} 1 & 0 & 0 \cr 0 & 1 & 0 \cr 0 & 0 & 0 \end{pmatrix},
  M_b = \begin{pmatrix} 0 & 1 & 0 \cr -1 & 0 & 0 \cr 0 & 0 & 0 \end{pmatrix},
  M_n = \begin{pmatrix} 0 & 0 & 0 \cr 0 & 0 & 0 \cr 0 & 0 & 1 \end{pmatrix}.
  \cr & &
 \label{correlation_domains_5}
\end{eqnarray}
With the notations
\begin{eqnarray}
N_{a,\ell} = R_\ell M_a R_\ell^{-1}, 
N_{b,\ell} = R_\ell M_b R_\ell^{-1},
N_{n,\ell} = R_\ell M_n R_\ell^{-1},\cr
\label{correlation_domains_4_1}
\end{eqnarray}
we get in the laboratory frame
\begin{eqnarray}
& &   \Lambda^{\gamma \gamma^\prime}_\ell({\bf q},\omega) = \label{correlation_domains_4_2} \\
& &   \Lambda^{aa}({\bf q},\omega) N^{\gamma \gamma^\prime}_{a,\ell} + \Lambda^{ab}({\bf q},\omega) N^{\gamma \gamma^\prime}_{b,\ell} + \Lambda^{nn}({\bf q},\omega) N^{\gamma \gamma^\prime}_{n,\ell}. 
\nonumber
\end{eqnarray}
The matrix $R_\ell$ is discussed in Appendix~\ref{matrices_R}. We shall denote as ${\tilde \lambda}_{Z;s,\ell} (\omega)$ the muon-spin lattice relaxation rate associated with the first two terms of Eq.~\ref{correlation_domains_4_bis}. This is the relaxation due to the phasons. 

As argued in Appendix~\ref{Long_pol_corr}, only ${\bf q}$ vectors very close to $\pm{\bf k}$ contribute to the relaxation of interest here. It is therefore a very good approximation to replace ${\mathcal A}^{\gamma\gamma^\prime}_s({\bf q})$ by ${\mathcal A}^{\gamma\gamma^\prime}_s(\pm{\bf k}_\ell)$ for $\Lambda^{aa}({\bf q},\omega)$ and $\Lambda^{ab}({\bf q},\omega)$, respectively (see Fig.~\ref{correlation}). We then derive
\begin{eqnarray}
{\tilde \lambda}_{Z;s,\ell} (\omega)  =  {\tilde \alpha}_{s,\ell} (\omega) T,
\label{lambdaZ_single_1_bis_bis}
\end{eqnarray}
with the notation
\begin{eqnarray}
{\tilde \alpha}_{s,\ell} (\omega)  =  \frac{ \pi {\mathcal D} k_{\rm B} k S}{2 \hbar c^{1/2}_\parallel} {\mathcal K}_{s,\ell}(\omega).
\label{lambdaZ_single_1_three}
\end{eqnarray}
Remarkably, calculating the required ${\bf q}$ integral (see e.g.\ Eq.~\ref{lambdaZ_single_1}) in cylindrical coordinates, an analytical expression for ${\mathcal K}_s(\omega)$ is derived:
\begin{eqnarray}
&& {\mathcal K}_{s,\ell}(\omega) = \label{lambda_k} \\
& & \frac{|\omega|}{16\pi k^2(c_\parallel c_\perp)^{1/2}}
\sum_{\gamma,\gamma^\prime} \left\{ \left[ {\mathcal A}^{\gamma\gamma^\prime}_s({\bf k}_\ell) + {\mathcal A}^{\gamma\gamma^\prime*}_s({\bf k}_\ell) \right] 
N_{a,\ell}^{\gamma\gamma^\prime} 
\vphantom{\frac{\tilde\Lambda_2^{\gamma\gamma^\prime}}{i}} \right. \nonumber \\ & & \hspace{25mm} \left. +
\left[{\mathcal A}^{\gamma\gamma^\prime}_s({\bf k}_\ell) - {\mathcal A}^{\gamma\gamma^\prime*}_s({\bf k}_\ell)\right]\frac{N_{b,\ell}^{\gamma\gamma^\prime}}{i}\right\}.
\nonumber 
\end{eqnarray}

Numerically it is found that the phason contribution to the relaxation is negligible compared to the contribution of the $nn$ correlation; see Appendix~\ref{Long_pol_rate}. Hence, $\mu$SR cannot provide information on phasons, at least for MnSi.

\subsection{Angular functions ${\mathcal V}^{\gamma \gamma^\prime}_s(\theta)$}
\label{functions}

We list the nine angular functions which appear in Eq.~\ref{Int_gamma_gamma_prime}:
\begin{eqnarray}
{\mathcal V}^{XX}_s (\theta)  & = &  8 \sin^4 \theta 
- 4 \left( \sin^2 \theta - 2 \frac{r_\mu H}{4\pi}\right) {\mathfrak C}^{XX}_s \cr & - &
16 \frac{r_\mu H}{4\pi}\sin^2 \theta + 16 \left(\frac{r_\mu H}{4\pi}\right)^2 \cr & + & \left({\mathfrak C}^{XX}_s\right)^2 + \left({\mathfrak C}^{XY}_s\right)^2, \cr 
{\mathcal V}^{Y Y}_s (\theta)  & = & 8 \sin^4\theta 
-4 \left(\sin^2 \theta - 2 \frac{r_\mu H}{4\pi}\right) {\mathfrak C}^{YY}_s \cr & - &
16 \frac{r_\mu H}{4\pi} \sin^2 \theta + 16 \left(\frac{r_\mu H}{4\pi}\right)^2 \cr & + &
\left({\mathfrak C}^{XY}_s\right)^2 + \left({\mathfrak C}^{YY}_s\right)^2, \cr
{\mathcal V}^{ZZ}_s (\theta)  & = &  16 \cos^2 \theta \sin^2\theta + \left({\mathfrak C}^{XZ}_s\right)^2 + \left({\mathfrak C}^{YZ}_s\right)^2, \cr
{\mathcal V}^{XY}_s (\theta)  & = & -4 \left(\sin^2 \theta - 2 \frac{r_\mu H}{4\pi}\right) {\mathfrak C}^{XY}_s \cr & + & {\mathfrak C}^{XY}_s\left( {\mathfrak C}^{XX}_s + {\mathfrak C}^{YY}_s\right),\cr
{\mathcal V}^{X Z}_s (\theta)  & = &  -2\left( \sin^2\theta -2 \frac{r_\mu H}{4\pi}\right){\mathfrak C}^{XZ}_s \cr & + &
{\mathfrak C}^{XX}_s {\mathfrak C}^{XZ}_s + {\mathfrak C}^{XY}_s {\mathfrak C}^{YZ}_s, \nonumber \\[1mm]
{\mathcal V}^{YX}_s (\theta) & = & {\mathcal V}^{XY}_s (\theta), \cr
{\mathcal V}^{Y Z}_s (\theta)  & = &  -2 \left( \sin^2\theta -2 \frac{r_\mu H}{4\pi}\right) {\mathfrak C}^{YZ}_s \cr & + &
{\mathfrak C}^{XY}_s {\mathfrak C}^{XZ}_s + {\mathfrak C}^{YY}_s {\mathfrak C}^{YZ}_s,\nonumber \\[1mm]
{\mathcal V}^{ZX}_s (\theta) & = & {\mathcal V}^{XZ}_s (\theta), \cr
{\mathcal V}^{ZY}_s (\theta) & = & {\mathcal V}^{YZ}_s (\theta),
\label{Int_gen}
\end{eqnarray}
with the definition ${\mathfrak C}^{\gamma\gamma^\prime}_s = \sum_{d} C^{\gamma\gamma^\prime}_{d,s}({\bf 0})$.

\subsection{Polarization function}
\label{polarization}
The polarization function resulting from Eq.~\ref{lambdaZ_dis_3} is a weighted sum of exponential functions. Numerically its amplitude at $t=0$ is found as expected to be 1/3. Moreover, for the constants $c_\parallel$ and $c_\perp$ introduced in Eq.~\ref{class_dynamics_22}, the functional form  of $P_Z^{\rm rel}(t,T)$ is very close to a stretched exponential shape $\exp[-(\lambda_Zt)^\beta]/3$, with $\beta \approx 0.87$. Now it must be realized that the overall relaxation is relatively weak, even when approaching 25~K where the experimental $\lambda_Z(T)$ starts to deviate from a linear temperature dependence. This weak relaxation combined with the intrinsically reduced amplitude of the signal associated with the 1/3 factor make the experimental observation of the deviation of $P_Z^{\rm rel}(t,T)$ from an exponential function nearly impossible. For the computations presented in this paper, $P_Z^{\rm rel}(t,T)$ was calculated at $T$ = 10~K in the range 0-6~$\mu$s where the data have most statistical weight and $b_\lambda$ was determined from a fit of $\exp(-b_\lambda T t)/3$ to $P_Z^{\rm rel}(t,T)$.

\subsection{Final comments}
\label{Com}

We provide two comments related to the theoretical description of $m(T)$ and $\lambda_Z(T)$.

First about the robustness of the $T^2$ and $T$ behaviours of $m(T)$ and $\lambda_Z(T)$ at low temperature, and its implication. The decay of $m(T)$ as $T^2$ directly reflects the cylindrical symmetry relative to the wavevector of the helimagnon dispersion relation. The linear thermal behaviour of $\lambda_Z (T)$ is due to the fact that the thermal energy is much greater than the energy of the helimagnons inducing the relaxation. The dependence of  $m(T)$ and $\lambda_Z(T)$ on the elastic constants $c_\parallel$ and $c_\perp$ is quite different. This is the reason for the successful determination of both constants.

The discussion of the spin-lattice relaxation takes into account the four muon magnetic sites, the four K-domains and the fact that three out the four muon sites probe a field distribution, rather than a single field. All of these features are important for a quantitative interpretation of the measurement. For instance, restricting the analysis to the $\ell =111$ K-domain results in a relaxation rate more than one order of magnitude smaller than observed. In fact, once the correlation functions are known, the account of the multiple muon sites and K-domains is just a geometrical problem, although tedious. We are in the favorable case where the muon position in the unit cell and its hyperfine coupling constant are known. Basically the only free physical parameters are the two elastic constants.

\bibliography{reference}

\begin{thebibliography}{69}%
\makeatletter
\providecommand \@ifxundefined [1]{%
 \@ifx{#1\undefined}
}%
\providecommand \@ifnum [1]{%
 \ifnum #1\expandafter \@firstoftwo
 \else \expandafter \@secondoftwo
 \fi
}%
\providecommand \@ifx [1]{%
 \ifx #1\expandafter \@firstoftwo
 \else \expandafter \@secondoftwo
 \fi
}%
\providecommand \natexlab [1]{#1}%
\providecommand \enquote  [1]{``#1''}%
\providecommand \bibnamefont  [1]{#1}%
\providecommand \bibfnamefont [1]{#1}%
\providecommand \citenamefont [1]{#1}%
\providecommand \href@noop [0]{\@secondoftwo}%
\providecommand \href [0]{\begingroup \@sanitize@url \@href}%
\providecommand \@href[1]{\@@startlink{#1}\@@href}%
\providecommand \@@href[1]{\endgroup#1\@@endlink}%
\providecommand \@sanitize@url [0]{\catcode `\\12\catcode `\$12\catcode
  `\&12\catcode `\#12\catcode `\^12\catcode `\_12\catcode `\%12\relax}%
\providecommand \@@startlink[1]{}%
\providecommand \@@endlink[0]{}%
\providecommand \url  [0]{\begingroup\@sanitize@url \@url }%
\providecommand \@url [1]{\endgroup\@href {#1}{\urlprefix }}%
\providecommand \urlprefix  [0]{URL }%
\providecommand \Eprint [0]{\href }%
\providecommand \doibase [0]{http://dx.doi.org/}%
\providecommand \selectlanguage [0]{\@gobble}%
\providecommand \bibinfo  [0]{\@secondoftwo}%
\providecommand \bibfield  [0]{\@secondoftwo}%
\providecommand \translation [1]{[#1]}%
\providecommand \BibitemOpen [0]{}%
\providecommand \bibitemStop [0]{}%
\providecommand \bibitemNoStop [0]{.\EOS\space}%
\providecommand \EOS [0]{\spacefactor3000\relax}%
\providecommand \BibitemShut  [1]{\csname bibitem#1\endcsname}%
\let\auto@bib@innerbib\@empty
\bibitem [{\citenamefont {Bor\'en}(1933)}]{Boren33}%
  \BibitemOpen
  \bibfield  {author} {\bibinfo {author} {\bibfnamefont {B.}~\bibnamefont
  {Bor\'en}},\ }\bibfield  {title} {\enquote {\bibinfo {title}
  {Roentgenuntersuchung der legierungen von silicium mit chrom, mangan, kobalt
  und nickel},}\ }\href@noop {} {\bibfield  {journal} {\bibinfo  {journal}
  {Arkiv f\"or Kemi, Mineralogi och Geologi}\ }\textbf {\bibinfo {volume}
  {11A}},\ \bibinfo {pages} {1} (\bibinfo {year} {1933})}\BibitemShut {NoStop}%
\bibitem [{\citenamefont {Dzyaloshinskii}(1958)}]{Dzyaloshinskii58}%
  \BibitemOpen
  \bibfield  {author} {\bibinfo {author} {\bibfnamefont {I~.~E.}\ \bibnamefont
  {Dzyaloshinskii}},\ }\bibfield  {title} {\enquote {\bibinfo {title} {A
  thermodynamic theory of "weak" ferromagnetism of antiferromagnetics},}\
  }\href {https://doi.org/10.1016/0022-3697(58)90076-3} {\bibfield  {journal}
  {\bibinfo  {journal} {J. Phys. Chem. Solids}\ }\textbf {\bibinfo {volume}
  {4}},\ \bibinfo {pages} {241} (\bibinfo {year} {1958})}\BibitemShut {NoStop}%
\bibitem [{\citenamefont {Moriya}(1960)}]{Moriya60}%
  \BibitemOpen
  \bibfield  {author} {\bibinfo {author} {\bibfnamefont {T\^oru}\ \bibnamefont
  {Moriya}},\ }\bibfield  {title} {\enquote {\bibinfo {title} {Anisotropic
  superexchange interaction and weak ferromagnetism},}\ }\href {\doibase
  10.1103/PhysRev.120.91} {\bibfield  {journal} {\bibinfo  {journal} {Phys.
  Rev.}\ }\textbf {\bibinfo {volume} {120}},\ \bibinfo {pages} {91--98}
  (\bibinfo {year} {1960})}\BibitemShut {NoStop}%
\bibitem [{\citenamefont {Ishikawa}\ \emph {et~al.}(1976)\citenamefont
  {Ishikawa}, \citenamefont {Tajima}, \citenamefont {Bloch},\ and\
  \citenamefont {Roth}}]{Ishikawa76}%
  \BibitemOpen
  \bibfield  {author} {\bibinfo {author} {\bibfnamefont {Y.}~\bibnamefont
  {Ishikawa}}, \bibinfo {author} {\bibfnamefont {K.}~\bibnamefont {Tajima}},
  \bibinfo {author} {\bibfnamefont {D.}~\bibnamefont {Bloch}}, \ and\ \bibinfo
  {author} {\bibfnamefont {M.}~\bibnamefont {Roth}},\ }\bibfield  {title}
  {\enquote {\bibinfo {title} {Helical spin structure in manganese silicide
  \uppercase{M}n\uppercase{S}i},}\ }\href {\doibase
  0.1016/0038-1098(76)90057-0} {\bibfield  {journal} {\bibinfo  {journal}
  {Solid State Commun.}\ }\textbf {\bibinfo {volume} {19}},\ \bibinfo {pages}
  {525} (\bibinfo {year} {1976})}\BibitemShut {NoStop}%
\bibitem [{\citenamefont {Bak}\ and\ \citenamefont {Jensen}(1980)}]{Bak80}%
  \BibitemOpen
  \bibfield  {author} {\bibinfo {author} {\bibfnamefont {P.}~\bibnamefont
  {Bak}}\ and\ \bibinfo {author} {\bibfnamefont {M.~H.}\ \bibnamefont
  {Jensen}},\ }\bibfield  {title} {\enquote {\bibinfo {title} {Theory of
  helical magnetic structures and phase transitions in
  \uppercase{M}n\uppercase{S}i and \uppercase{F}e\uppercase{G}e},}\ }\href
  {\doibase 10.1088/0022-3719/13/31/002} {\bibfield  {journal} {\bibinfo
  {journal} {J. Phys. C: Solid State Phys.}\ }\textbf {\bibinfo {volume}
  {13}},\ \bibinfo {pages} {L881} (\bibinfo {year} {1980})}\BibitemShut
  {NoStop}%
\bibitem [{\citenamefont {Nakanishi}\ \emph {et~al.}(1980)\citenamefont
  {Nakanishi}, \citenamefont {Yanase}, \citenamefont {Hasegawa},\ and\
  \citenamefont {Kataoka}}]{Nakanishi80}%
  \BibitemOpen
  \bibfield  {author} {\bibinfo {author} {\bibfnamefont {O.}~\bibnamefont
  {Nakanishi}}, \bibinfo {author} {\bibfnamefont {A.}~\bibnamefont {Yanase}},
  \bibinfo {author} {\bibfnamefont {A.}~\bibnamefont {Hasegawa}}, \ and\
  \bibinfo {author} {\bibfnamefont {M.}~\bibnamefont {Kataoka}},\ }\bibfield
  {title} {\enquote {\bibinfo {title} {The origin of the helical spin density
  wave in \uppercase{M}n\uppercase{S}i},}\ }\href {\doibase
  10.1016/0038-1098(80)91004-2} {\bibfield  {journal} {\bibinfo  {journal}
  {Solid State Commun.}\ }\textbf {\bibinfo {volume} {35}},\ \bibinfo {pages}
  {995} (\bibinfo {year} {1980})}\BibitemShut {NoStop}%
\bibitem [{\citenamefont {Lonzarich}\ and\ \citenamefont
  {Taillefer}(1985)}]{Lonzarich85}%
  \BibitemOpen
  \bibfield  {author} {\bibinfo {author} {\bibfnamefont {G~G}\ \bibnamefont
  {Lonzarich}}\ and\ \bibinfo {author} {\bibfnamefont {L}~\bibnamefont
  {Taillefer}},\ }\bibfield  {title} {\enquote {\bibinfo {title} {Effect of
  spin fluctuations on the magnetic equation of state of ferromagnetic or
  nearly ferromagnetic metals},}\ }\href
  {https://doi.org/10.1088/0022-3719/18/22/017} {\bibfield  {journal} {\bibinfo
   {journal} {J. Phys. C: Solid State Phys.}\ }\textbf {\bibinfo {volume}
  {18}},\ \bibinfo {pages} {4339} (\bibinfo {year} {1985})}\BibitemShut
  {NoStop}%
\bibitem [{\citenamefont {Moriya}(1985)}]{Moriya85}%
  \BibitemOpen
  \bibfield  {author} {\bibinfo {author} {\bibfnamefont {T.}~\bibnamefont
  {Moriya}},\ }\href@noop {} {\emph {\bibinfo {title} {Spin fluctuations in
  itinerant electron magnetism}}}\ (\bibinfo  {publisher} {Springer},\ \bibinfo
  {address} {Berlin},\ \bibinfo {year} {1985})\BibitemShut {NoStop}%
\bibitem [{\citenamefont {Kakehashi}(2013)}]{Kakehashi13}%
  \BibitemOpen
  \bibfield  {author} {\bibinfo {author} {\bibfnamefont {Y.}~\bibnamefont
  {Kakehashi}},\ }\href@noop {} {\emph {\bibinfo {title} {Modern theory of
  magnetism in metals and alloys}}}\ (\bibinfo  {publisher} {Springer},\
  \bibinfo {address} {Berlin},\ \bibinfo {year} {2013})\BibitemShut {NoStop}%
\bibitem [{\citenamefont {Pfleiderer}\ \emph {et~al.}(2001)\citenamefont
  {Pfleiderer}, \citenamefont {Julian},\ and\ \citenamefont
  {Lonzarich}}]{Pfleiderer01}%
  \BibitemOpen
  \bibfield  {author} {\bibinfo {author} {\bibfnamefont {C.}~\bibnamefont
  {Pfleiderer}}, \bibinfo {author} {\bibfnamefont {S.~R.}\ \bibnamefont
  {Julian}}, \ and\ \bibinfo {author} {\bibfnamefont {G.~G.}\ \bibnamefont
  {Lonzarich}},\ }\bibfield  {title} {\enquote {\bibinfo {title}
  {Non-\uppercase{F}ermi liquid nature of the normal state of
  itinerant-electron ferromagnets},}\ }\href {\doibase 10.1038/35106527}
  {\bibfield  {journal} {\bibinfo  {journal} {Nature}\ }\textbf {\bibinfo
  {volume} {414}},\ \bibinfo {pages} {427} (\bibinfo {year}
  {2001})}\BibitemShut {NoStop}%
\bibitem [{\citenamefont {Pfleiderer}\ \emph {et~al.}(2004)\citenamefont
  {Pfleiderer}, \citenamefont {Reznik}, \citenamefont {Pintschovius},
  \citenamefont {{von L\"ohneysen}}, \citenamefont {Garst},\ and\ \citenamefont
  {Rosch}}]{Pfleiderer04}%
  \BibitemOpen
  \bibfield  {author} {\bibinfo {author} {\bibfnamefont {C.}~\bibnamefont
  {Pfleiderer}}, \bibinfo {author} {\bibfnamefont {D.}~\bibnamefont {Reznik}},
  \bibinfo {author} {\bibfnamefont {L.}~\bibnamefont {Pintschovius}}, \bibinfo
  {author} {\bibfnamefont {H.}~\bibnamefont {{von L\"ohneysen}}}, \bibinfo
  {author} {\bibfnamefont {M.}~\bibnamefont {Garst}}, \ and\ \bibinfo {author}
  {\bibfnamefont {A.}~\bibnamefont {Rosch}},\ }\bibfield  {title} {\enquote
  {\bibinfo {title} {Partial order in the non-\uppercase{F}ermi-liquid phase of
  \uppercase{M}n\uppercase{S}i},}\ }\href {\doibase 10.1038/nature02232}
  {\bibfield  {journal} {\bibinfo  {journal} {Nature}\ }\textbf {\bibinfo
  {volume} {427}},\ \bibinfo {pages} {227} (\bibinfo {year}
  {2004})}\BibitemShut {NoStop}%
\bibitem [{\citenamefont {Bannenberg}\ \emph {et~al.}(2019)\citenamefont
  {Bannenberg}, \citenamefont {Sadykov}, \citenamefont {Dalgliesh},
  \citenamefont {Goodway}, \citenamefont {Schlagel}, \citenamefont {Lograsso},
  \citenamefont {Falus}, \citenamefont {Leli\`evre-Berna}, \citenamefont
  {Leonov},\ and\ \citenamefont {Pappas}}]{Bannenberg19}%
  \BibitemOpen
  \bibfield  {author} {\bibinfo {author} {\bibfnamefont {L.~J.}\ \bibnamefont
  {Bannenberg}}, \bibinfo {author} {\bibfnamefont {R.}~\bibnamefont {Sadykov}},
  \bibinfo {author} {\bibfnamefont {R.~M.}\ \bibnamefont {Dalgliesh}}, \bibinfo
  {author} {\bibfnamefont {C.}~\bibnamefont {Goodway}}, \bibinfo {author}
  {\bibfnamefont {D.~L.}\ \bibnamefont {Schlagel}}, \bibinfo {author}
  {\bibfnamefont {T.~A.}\ \bibnamefont {Lograsso}}, \bibinfo {author}
  {\bibfnamefont {P.}~\bibnamefont {Falus}}, \bibinfo {author} {\bibfnamefont
  {E.}~\bibnamefont {Leli\`evre-Berna}}, \bibinfo {author} {\bibfnamefont
  {A.~O.}\ \bibnamefont {Leonov}}, \ and\ \bibinfo {author} {\bibfnamefont
  {C.}~\bibnamefont {Pappas}},\ }\bibfield  {title} {\enquote {\bibinfo {title}
  {Skyrmions and spirals in \uppercase{M}n\uppercase{S}i under hydrostatic
  pressure},}\ }\href {\doibase 10.1103/PhysRevB.100.054447} {\bibfield
  {journal} {\bibinfo  {journal} {Phys. Rev. B}\ }\textbf {\bibinfo {volume}
  {100}},\ \bibinfo {pages} {054447} (\bibinfo {year} {2019})}\BibitemShut
  {NoStop}%
\bibitem [{\citenamefont {M\"uhlbauer}\ \emph {et~al.}(2009)\citenamefont
  {M\"uhlbauer}, \citenamefont {Binz}, \citenamefont {Jonietz}, \citenamefont
  {Pfleiderer}, \citenamefont {Rosch}, \citenamefont {Neubauer}, \citenamefont
  {Georgii},\ and\ \citenamefont {B\"oni}}]{Muhlbauer09a}%
  \BibitemOpen
  \bibfield  {author} {\bibinfo {author} {\bibfnamefont {S.}~\bibnamefont
  {M\"uhlbauer}}, \bibinfo {author} {\bibfnamefont {B.}~\bibnamefont {Binz}},
  \bibinfo {author} {\bibfnamefont {F.}~\bibnamefont {Jonietz}}, \bibinfo
  {author} {\bibfnamefont {C.}~\bibnamefont {Pfleiderer}}, \bibinfo {author}
  {\bibfnamefont {A.}~\bibnamefont {Rosch}}, \bibinfo {author} {\bibfnamefont
  {A.}~\bibnamefont {Neubauer}}, \bibinfo {author} {\bibfnamefont
  {R.}~\bibnamefont {Georgii}}, \ and\ \bibinfo {author} {\bibfnamefont
  {P.}~\bibnamefont {B\"oni}},\ }\bibfield  {title} {\enquote {\bibinfo {title}
  {Skyrmion lattice in a chiral magnet},}\ }\href {\doibase
  10.1126/science.1166767} {\bibfield  {journal} {\bibinfo  {journal}
  {Science}\ }\textbf {\bibinfo {volume} {323}},\ \bibinfo {pages} {915}
  (\bibinfo {year} {2009})},\ \bibinfo {note} {see also erratum to this
  article}\BibitemShut {NoStop}%
\bibitem [{\citenamefont {Dalmas~de R\'eotier}\ \emph
  {et~al.}(2016)\citenamefont {Dalmas~de R\'eotier}, \citenamefont
  {Maisuradze}, \citenamefont {Yaouanc}, \citenamefont {Roessli}, \citenamefont
  {Amato}, \citenamefont {Andreica},\ and\ \citenamefont
  {Lapertot}}]{Dalmas16}%
  \BibitemOpen
  \bibfield  {author} {\bibinfo {author} {\bibfnamefont {P.}~\bibnamefont
  {Dalmas~de R\'eotier}}, \bibinfo {author} {\bibfnamefont {A.}~\bibnamefont
  {Maisuradze}}, \bibinfo {author} {\bibfnamefont {A.}~\bibnamefont {Yaouanc}},
  \bibinfo {author} {\bibfnamefont {B.}~\bibnamefont {Roessli}}, \bibinfo
  {author} {\bibfnamefont {A.}~\bibnamefont {Amato}}, \bibinfo {author}
  {\bibfnamefont {D.}~\bibnamefont {Andreica}}, \ and\ \bibinfo {author}
  {\bibfnamefont {G.}~\bibnamefont {Lapertot}},\ }\bibfield  {title} {\enquote
  {\bibinfo {title} {Determination of the zero-field magnetic structure of the
  helimagnet \uppercase{M}n\uppercase{S}i at low temperature},}\ }\href
  {\doibase 10.1103/PhysRevB.93.144419} {\bibfield  {journal} {\bibinfo
  {journal} {Phys. Rev. B}\ }\textbf {\bibinfo {volume} {93}},\ \bibinfo
  {pages} {144419} (\bibinfo {year} {2016})}\BibitemShut {NoStop}%
\bibitem [{\citenamefont {Dalmas~de R\'eotier}\ \emph
  {et~al.}(2017)\citenamefont {Dalmas~de R\'eotier}, \citenamefont
  {Maisuradze}, \citenamefont {Yaouanc}, \citenamefont {Roessli}, \citenamefont
  {Amato}, \citenamefont {Andreica},\ and\ \citenamefont
  {Lapertot}}]{Dalmas17}%
  \BibitemOpen
  \bibfield  {author} {\bibinfo {author} {\bibfnamefont {P.}~\bibnamefont
  {Dalmas~de R\'eotier}}, \bibinfo {author} {\bibfnamefont {A.}~\bibnamefont
  {Maisuradze}}, \bibinfo {author} {\bibfnamefont {A.}~\bibnamefont {Yaouanc}},
  \bibinfo {author} {\bibfnamefont {B.}~\bibnamefont {Roessli}}, \bibinfo
  {author} {\bibfnamefont {A.}~\bibnamefont {Amato}}, \bibinfo {author}
  {\bibfnamefont {D.}~\bibnamefont {Andreica}}, \ and\ \bibinfo {author}
  {\bibfnamefont {G.}~\bibnamefont {Lapertot}},\ }\bibfield  {title} {\enquote
  {\bibinfo {title} {Unconventional magnetic order in the conical state of
  \uppercase{M}n\uppercase{S}i},}\ }\href {\doibase 10.1103/PhysRevB.95.180403}
  {\bibfield  {journal} {\bibinfo  {journal} {Phys. Rev. B}\ }\textbf {\bibinfo
  {volume} {95}},\ \bibinfo {pages} {180403(R)} (\bibinfo {year}
  {2017})}\BibitemShut {NoStop}%
\bibitem [{\citenamefont {{Dalmas de R\'eotier}}\ \emph
  {et~al.}(2018)\citenamefont {{Dalmas de R\'eotier}}, \citenamefont {Yaouanc},
  \citenamefont {Amato}, \citenamefont {Maisuradze}, \citenamefont {Andreica},
  \citenamefont {Roessli}, \citenamefont {Goko}, \citenamefont {Scheuermann},\
  and\ \citenamefont {Lapertot}}]{Dalmas18}%
  \BibitemOpen
  \bibfield  {author} {\bibinfo {author} {\bibfnamefont {P.}~\bibnamefont
  {{Dalmas de R\'eotier}}}, \bibinfo {author} {\bibfnamefont {A.}~\bibnamefont
  {Yaouanc}}, \bibinfo {author} {\bibfnamefont {A.}~\bibnamefont {Amato}},
  \bibinfo {author} {\bibfnamefont {A}~\bibnamefont {Maisuradze}}, \bibinfo
  {author} {\bibfnamefont {D.}~\bibnamefont {Andreica}}, \bibinfo {author}
  {\bibfnamefont {B.}~\bibnamefont {Roessli}}, \bibinfo {author} {\bibfnamefont
  {T.}~\bibnamefont {Goko}}, \bibinfo {author} {\bibfnamefont {R.}~\bibnamefont
  {Scheuermann}}, \ and\ \bibinfo {author} {\bibfnamefont {G.}~\bibnamefont
  {Lapertot}},\ }\bibfield  {title} {\enquote {\bibinfo {title} {On the
  robustness of the \uppercase{M}n\uppercase{S}i magnetic structure determined
  by muon spin rotation},}\ }\href {\doibase 10.3390/qubs2030019} {\bibfield
  {journal} {\bibinfo  {journal} {Quantum Beam Sci.}\ }\textbf {\bibinfo
  {volume} {2}},\ \bibinfo {pages} {19} (\bibinfo {year} {2018})}\BibitemShut
  {NoStop}%
\bibitem [{\citenamefont {Thessieu}\ \emph {et~al.}(1998)\citenamefont
  {Thessieu}, \citenamefont {Kamishima}, \citenamefont {Goto},\ and\
  \citenamefont {Lapertot}}]{Thessieu98a}%
  \BibitemOpen
  \bibfield  {author} {\bibinfo {author} {\bibfnamefont {C.}~\bibnamefont
  {Thessieu}}, \bibinfo {author} {\bibfnamefont {K.}~\bibnamefont {Kamishima}},
  \bibinfo {author} {\bibfnamefont {T.}~\bibnamefont {Goto}}, \ and\ \bibinfo
  {author} {\bibfnamefont {G.}~\bibnamefont {Lapertot}},\ }\bibfield  {title}
  {\enquote {\bibinfo {title} {Magnetization under high pressure in
  \uppercase{M}n\uppercase{S}i},}\ }\href
  {https://doi.org/10.1143/JPSJ.67.3605} {\bibfield  {journal} {\bibinfo
  {journal} {J. Phys. Soc. Jpn.}\ }\textbf {\bibinfo {volume} {67}},\ \bibinfo
  {pages} {3605} (\bibinfo {year} {1998})}\BibitemShut {NoStop}%
\bibitem [{\citenamefont {Yaouanc}\ \emph {et~al.}(2005)\citenamefont
  {Yaouanc}, \citenamefont {{Dalmas de R\'eotier}}, \citenamefont {Gubbens},
  \citenamefont {Sakarya}, \citenamefont {Lapertot}, \citenamefont {Hillier},\
  and\ \citenamefont {King}}]{Yaouanc05}%
  \BibitemOpen
  \bibfield  {author} {\bibinfo {author} {\bibfnamefont {A.}~\bibnamefont
  {Yaouanc}}, \bibinfo {author} {\bibfnamefont {P.}~\bibnamefont {{Dalmas de
  R\'eotier}}}, \bibinfo {author} {\bibfnamefont {P.~C.~M.}\ \bibnamefont
  {Gubbens}}, \bibinfo {author} {\bibfnamefont {S.}~\bibnamefont {Sakarya}},
  \bibinfo {author} {\bibfnamefont {G.}~\bibnamefont {Lapertot}}, \bibinfo
  {author} {\bibfnamefont {A.~D.}\ \bibnamefont {Hillier}}, \ and\ \bibinfo
  {author} {\bibfnamefont {P.~J.~C.}\ \bibnamefont {King}},\ }\bibfield
  {title} {\enquote {\bibinfo {title} {Testing the self-consistent
  renormalization theory for the description of the spin-fluctuation modes of
  \uppercase{M}n\uppercase{S}i at ambient pressure},}\ }\href {\doibase
  10.1088/0953-8984/17/13/L01} {\bibfield  {journal} {\bibinfo  {journal} {J.
  Phys.: Condens. Matter}\ }\textbf {\bibinfo {volume} {17}},\ \bibinfo {pages}
  {L129} (\bibinfo {year} {2005})}\BibitemShut {NoStop}%
\bibitem [{\citenamefont {Moriya}\ and\ \citenamefont {Ueda}(1974)}]{Moriya74}%
  \BibitemOpen
  \bibfield  {author} {\bibinfo {author} {\bibfnamefont {T\^oru}\ \bibnamefont
  {Moriya}}\ and\ \bibinfo {author} {\bibfnamefont {Kazuo}\ \bibnamefont
  {Ueda}},\ }\bibfield  {title} {\enquote {\bibinfo {title} {Nuclear magnetic
  relaxation in weakly ferro-and antiferromagnetic metals},}\ }\href {\doibase
  https://doi.org/10.1016/0038-1098(74)90733-9} {\bibfield  {journal} {\bibinfo
   {journal} {Solid State Communications}\ }\textbf {\bibinfo {volume} {15}},\
  \bibinfo {pages} {169 -- 172} (\bibinfo {year} {1974})}\BibitemShut {NoStop}%
\bibitem [{\citenamefont {Yaouanc}\ and\ \citenamefont {{Dalmas de
  R\'eotier}}(2011)}]{Yaouanc11}%
  \BibitemOpen
  \bibfield  {author} {\bibinfo {author} {\bibfnamefont {A.}~\bibnamefont
  {Yaouanc}}\ and\ \bibinfo {author} {\bibfnamefont {P.}~\bibnamefont {{Dalmas
  de R\'eotier}}},\ }\href@noop {} {\emph {\bibinfo {title} {Muon Spin
  Rotation, Relaxation, and Resonance: Applications to Condensed Matter}}}\
  (\bibinfo  {publisher} {Oxford University Press},\ \bibinfo {address}
  {Oxford},\ \bibinfo {year} {2011})\BibitemShut {NoStop}%
\bibitem [{\citenamefont {Amato}\ \emph {et~al.}(2017)\citenamefont {Amato},
  \citenamefont {Luetkens}, \citenamefont {Sedlak}, \citenamefont {Stoykov},
  \citenamefont {Scheuermann}, \citenamefont {Elender}, \citenamefont
  {Raselli},\ and\ \citenamefont {Graf}}]{Amato17}%
  \BibitemOpen
  \bibfield  {author} {\bibinfo {author} {\bibfnamefont {A.}~\bibnamefont
  {Amato}}, \bibinfo {author} {\bibfnamefont {H.}~\bibnamefont {Luetkens}},
  \bibinfo {author} {\bibfnamefont {K.}~\bibnamefont {Sedlak}}, \bibinfo
  {author} {\bibfnamefont {A.}~\bibnamefont {Stoykov}}, \bibinfo {author}
  {\bibfnamefont {R.}~\bibnamefont {Scheuermann}}, \bibinfo {author}
  {\bibfnamefont {M.}~\bibnamefont {Elender}}, \bibinfo {author} {\bibfnamefont
  {A.}~\bibnamefont {Raselli}}, \ and\ \bibinfo {author} {\bibfnamefont
  {D.}~\bibnamefont {Graf}},\ }\bibfield  {title} {\enquote {\bibinfo {title}
  {The new versatile general purpose surface-muon instrument (\uppercase{GPS})
  based on silicon photomultipliers for $\mu$\uppercase{SR} measurements on a
  continuous-wave beam},}\ }\href {https://doi.org/10.1063/1.4986045}
  {\bibfield  {journal} {\bibinfo  {journal} {Rev. Sci. Instrum.}\ }\textbf
  {\bibinfo {volume} {88}},\ \bibinfo {pages} {093301} (\bibinfo {year}
  {2017})}\BibitemShut {NoStop}%
\bibitem [{\citenamefont {Yaouanc}\ \emph {et~al.}(2017)\citenamefont
  {Yaouanc}, \citenamefont {Dalmas~de R\'eotier}, \citenamefont {Maisuradze},\
  and\ \citenamefont {Roessli}}]{Yaouanc17}%
  \BibitemOpen
  \bibfield  {author} {\bibinfo {author} {\bibfnamefont {A.}~\bibnamefont
  {Yaouanc}}, \bibinfo {author} {\bibfnamefont {P.}~\bibnamefont {Dalmas~de
  R\'eotier}}, \bibinfo {author} {\bibfnamefont {A.}~\bibnamefont
  {Maisuradze}}, \ and\ \bibinfo {author} {\bibfnamefont {B.}~\bibnamefont
  {Roessli}},\ }\bibfield  {title} {\enquote {\bibinfo {title} {Magnetic
  structure of the \uppercase{M}n\uppercase{G}e helimagnet and representation
  analysis},}\ }\href {\doibase 10.1103/PhysRevB.95.174422} {\bibfield
  {journal} {\bibinfo  {journal} {Phys. Rev. B}\ }\textbf {\bibinfo {volume}
  {95}},\ \bibinfo {pages} {174422} (\bibinfo {year} {2017})}\BibitemShut
  {NoStop}%
\bibitem [{\citenamefont {Maisuradze}\ \emph {et~al.}(2018)\citenamefont
  {Maisuradze}, \citenamefont {Yaouanc},\ and\ \citenamefont {{Dalmas de
  R\'eotier}}}]{Maisuradze18}%
  \BibitemOpen
  \bibfield  {author} {\bibinfo {author} {\bibfnamefont {A.}~\bibnamefont
  {Maisuradze}}, \bibinfo {author} {\bibfnamefont {A.}~\bibnamefont {Yaouanc}},
  \ and\ \bibinfo {author} {\bibfnamefont {P.}~\bibnamefont {{Dalmas de
  R\'eotier}}},\ }\bibfield  {title} {\enquote {\bibinfo {title} {Reverse
  \uppercase{M}onte \uppercase{C}arlo algorithm and maximum entropy principle
  for the analysis of positive muon spin rotation and relaxation spectra},}\
  }\href {\doibase 10.7566/JPSCP.21.011053} {\bibfield  {journal} {\bibinfo
  {journal} {JPS Conf. Proc.}\ }\textbf {\bibinfo {volume} {21}},\ \bibinfo
  {pages} {011053} (\bibinfo {year} {2018})}\BibitemShut {NoStop}%
\bibitem [{\citenamefont {Amato}\ \emph {et~al.}(2014)\citenamefont {Amato},
  \citenamefont {{Dalmas de R\'eotier}}, \citenamefont {Andreica},
  \citenamefont {Yaouanc}, \citenamefont {Suter}, \citenamefont {Lapertot},
  \citenamefont {Pop}, \citenamefont {Morenzoni}, \citenamefont {Bonf\`a},
  \citenamefont {Bernardini},\ and\ \citenamefont {De~Renzi}}]{Amato14}%
  \BibitemOpen
  \bibfield  {author} {\bibinfo {author} {\bibfnamefont {A.}~\bibnamefont
  {Amato}}, \bibinfo {author} {\bibfnamefont {P.}~\bibnamefont {{Dalmas de
  R\'eotier}}}, \bibinfo {author} {\bibfnamefont {D.}~\bibnamefont {Andreica}},
  \bibinfo {author} {\bibfnamefont {A.}~\bibnamefont {Yaouanc}}, \bibinfo
  {author} {\bibfnamefont {A.}~\bibnamefont {Suter}}, \bibinfo {author}
  {\bibfnamefont {G.}~\bibnamefont {Lapertot}}, \bibinfo {author}
  {\bibfnamefont {I.~M.}\ \bibnamefont {Pop}}, \bibinfo {author} {\bibfnamefont
  {E.}~\bibnamefont {Morenzoni}}, \bibinfo {author} {\bibfnamefont
  {P.}~\bibnamefont {Bonf\`a}}, \bibinfo {author} {\bibfnamefont
  {F.}~\bibnamefont {Bernardini}}, \ and\ \bibinfo {author} {\bibfnamefont
  {R.}~\bibnamefont {De~Renzi}},\ }\bibfield  {title} {\enquote {\bibinfo
  {title} {Understanding the $\mu$\uppercase{SR} spectra of
  \uppercase{M}n\uppercase{S}i without magnetic polarons},}\ }\href {\doibase
  10.1103/PhysRevB.89.184425} {\bibfield  {journal} {\bibinfo  {journal} {Phys.
  Rev. B}\ }\textbf {\bibinfo {volume} {89}},\ \bibinfo {pages} {184425}
  (\bibinfo {year} {2014})}\BibitemShut {NoStop}%
\bibitem [{Note1()}]{Note1}%
  \BibitemOpen
  \bibinfo {note} {The modulus of the propagation wavevector increases from $k
  \simeq 0.35 \protect \tmspace +\thinmuskip {.1667em} {\protect \rm nm}^{-1}$
  for $T \rightarrow 0$ to $k \simeq 0.38 \protect \tmspace +\thinmuskip
  {.1667em} {\protect \rm nm}^{-1}$ for $T \rightarrow T_c$ \cite
  {Ishikawa76,Grigoriev06,Janoschek13}. This variation has a negligible
  influence on $D_{\protect \rm osc}$; see Fig.~1(a) of Ref.~\protect
  \rev@citealp {Dalmas18}}\BibitemShut {NoStop}%
\bibitem [{\citenamefont {Nagamiya}(1967)}]{Nagamiya67}%
  \BibitemOpen
  \bibfield  {author} {\bibinfo {author} {\bibfnamefont {T.}~\bibnamefont
  {Nagamiya}},\ }\bibfield  {title} {\enquote {\bibinfo {title} {Helical spin
  ordering --- 1 \uppercase{T}heory of helical spin configurations},}\ }in\
  \href@noop {} {\emph {\bibinfo {booktitle} {Solid State Physics, Advances in
  Research and Applications}}},\ Vol.~\bibinfo {volume} {20},\ \bibinfo
  {editor} {edited by\ \bibinfo {editor} {\bibfnamefont {D.~T.~F.}\
  \bibnamefont {Seitz}}\ and\ \bibinfo {editor} {\bibfnamefont
  {H.}~\bibnamefont {Ehrenreich}}}\ (\bibinfo  {publisher} {Academic Press},\
  \bibinfo {address} {New York},\ \bibinfo {year} {1967})\BibitemShut {NoStop}%
\bibitem [{\citenamefont {{Dalmas de R\'eotier}}\ and\ \citenamefont
  {Yaouanc}(1997)}]{Dalmas97}%
  \BibitemOpen
  \bibfield  {author} {\bibinfo {author} {\bibfnamefont {P.}~\bibnamefont
  {{Dalmas de R\'eotier}}}\ and\ \bibinfo {author} {\bibfnamefont
  {A.}~\bibnamefont {Yaouanc}},\ }\bibfield  {title} {\enquote {\bibinfo
  {title} {Muon spin rotation and relaxation in magnetic materials},}\ }\href
  {\doibase 10.1088/0953-8984/9/43/002} {\bibfield  {journal} {\bibinfo
  {journal} {J. Phys.: Condens. Matter}\ }\textbf {\bibinfo {volume} {9}},\
  \bibinfo {pages} {9113} (\bibinfo {year} {1997})}\BibitemShut {NoStop}%
\bibitem [{\citenamefont {{Dalmas de R\'eotier}}\ \emph
  {et~al.}(2004)\citenamefont {{Dalmas de R\'eotier}}, \citenamefont
  {Gubbens},\ and\ \citenamefont {Yaouanc}}]{Dalmas04}%
  \BibitemOpen
  \bibfield  {author} {\bibinfo {author} {\bibfnamefont {P.}~\bibnamefont
  {{Dalmas de R\'eotier}}}, \bibinfo {author} {\bibfnamefont {P.~C.~M.}\
  \bibnamefont {Gubbens}}, \ and\ \bibinfo {author} {\bibfnamefont
  {A.}~\bibnamefont {Yaouanc}},\ }\bibfield  {title} {\enquote {\bibinfo
  {title} {Probing magnetic excitations, fluctuations and correlations lengths
  by muon spin relaxation and rotation techniques},}\ }\href {\doibase
  10.1088/0953-8984/16/40/014} {\bibfield  {journal} {\bibinfo  {journal} {J.
  Phys.: Condens. Matter}\ }\textbf {\bibinfo {volume} {16}},\ \bibinfo {pages}
  {S4687} (\bibinfo {year} {2004})}\BibitemShut {NoStop}%
\bibitem [{\citenamefont {Hu}(2018)}]{Hu18}%
  \BibitemOpen
  \bibfield  {author} {\bibinfo {author} {\bibfnamefont {Yangfan}\ \bibnamefont
  {Hu}},\ }\bibfield  {title} {\enquote {\bibinfo {title} {Direction-dependent
  stability of skyrmion lattice in helimagnets induced by exchange
  anisotropy},}\ }\href {https://doi.org/10.1016/j.jmmm.2017.12.046} {\bibfield
   {journal} {\bibinfo  {journal} {J. Magn. Magn. Mater.}\ }\textbf {\bibinfo
  {volume} {455}},\ \bibinfo {pages} {54} (\bibinfo {year} {2018})}\BibitemShut
  {NoStop}%
\bibitem [{\citenamefont {Maleyev}(2006)}]{Maleyev06}%
  \BibitemOpen
  \bibfield  {author} {\bibinfo {author} {\bibfnamefont {S.~V.}\ \bibnamefont
  {Maleyev}},\ }\bibfield  {title} {\enquote {\bibinfo {title} {Cubic magnets
  with \uppercase{D}zyaloshinskii-\uppercase{M}oriya interaction at low
  temperature},}\ }\href {\doibase 10.1103/PhysRevB.73.174402} {\bibfield
  {journal} {\bibinfo  {journal} {Phys. Rev. B}\ }\textbf {\bibinfo {volume}
  {73}},\ \bibinfo {pages} {174402} (\bibinfo {year} {2006})}\BibitemShut
  {NoStop}%
\bibitem [{\citenamefont {Belitz}\ \emph {et~al.}(2006)\citenamefont {Belitz},
  \citenamefont {Kirkpatrick},\ and\ \citenamefont {Rosch}}]{Belitz06}%
  \BibitemOpen
  \bibfield  {author} {\bibinfo {author} {\bibfnamefont {D.}~\bibnamefont
  {Belitz}}, \bibinfo {author} {\bibfnamefont {T.~R.}\ \bibnamefont
  {Kirkpatrick}}, \ and\ \bibinfo {author} {\bibfnamefont {A.}~\bibnamefont
  {Rosch}},\ }\bibfield  {title} {\enquote {\bibinfo {title} {Theory of
  helimagnons in itinerant quantum systems},}\ }\href {\doibase
  10.1103/PhysRevB.73.054431} {\bibfield  {journal} {\bibinfo  {journal} {Phys.
  Rev. B}\ }\textbf {\bibinfo {volume} {73}},\ \bibinfo {pages} {054431}
  (\bibinfo {year} {2006})}\BibitemShut {NoStop}%
\bibitem [{\citenamefont {Ho}\ \emph {et~al.}(2010)\citenamefont {Ho},
  \citenamefont {Kirkpatrick}, \citenamefont {Sang},\ and\ \citenamefont
  {Belitz}}]{Ho10}%
  \BibitemOpen
  \bibfield  {author} {\bibinfo {author} {\bibfnamefont {Kwan-yuet}\
  \bibnamefont {Ho}}, \bibinfo {author} {\bibfnamefont {T.~R.}\ \bibnamefont
  {Kirkpatrick}}, \bibinfo {author} {\bibfnamefont {Yan}\ \bibnamefont {Sang}},
  \ and\ \bibinfo {author} {\bibfnamefont {D.}~\bibnamefont {Belitz}},\
  }\bibfield  {title} {\enquote {\bibinfo {title} {Ordered phases of itinerant
  \uppercase{D}zyaloshinsky-\uppercase{M}oriya magnets and their electronic
  properties},}\ }\href {\doibase 10.1103/PhysRevB.82.134427} {\bibfield
  {journal} {\bibinfo  {journal} {Phys. Rev. B}\ }\textbf {\bibinfo {volume}
  {82}},\ \bibinfo {pages} {134427} (\bibinfo {year} {2010})}\BibitemShut
  {NoStop}%
\bibitem [{\citenamefont {Date}\ \emph {et~al.}(1977)\citenamefont {Date},
  \citenamefont {Okuda},\ and\ \citenamefont {Kadowaki}}]{Date77}%
  \BibitemOpen
  \bibfield  {author} {\bibinfo {author} {\bibfnamefont {Muneyuki}\
  \bibnamefont {Date}}, \bibinfo {author} {\bibfnamefont {Kiichi}\ \bibnamefont
  {Okuda}}, \ and\ \bibinfo {author} {\bibfnamefont {Kazuo}\ \bibnamefont
  {Kadowaki}},\ }\bibfield  {title} {\enquote {\bibinfo {title} {Electron spin
  resonance in the itinerant-electron helical magnet
  \uppercase{M}n\uppercase{S}i},}\ }\href
  {https://doi.org/10.1143/JPSJ.42.1555} {\bibfield  {journal} {\bibinfo
  {journal} {J. Phys. Soc. Jpn.}\ }\textbf {\bibinfo {volume} {42}},\ \bibinfo
  {pages} {1555} (\bibinfo {year} {1977})}\BibitemShut {NoStop}%
\bibitem [{\citenamefont {Demishev}\ \emph {et~al.}(2011)\citenamefont
  {Demishev}, \citenamefont {Semeno}, \citenamefont {Bogach}, \citenamefont
  {Glushkov}, \citenamefont {Sluchanko}, \citenamefont {Samarin},\ and\
  \citenamefont {Chernobrovkin}}]{Demishev11}%
  \BibitemOpen
  \bibfield  {author} {\bibinfo {author} {\bibfnamefont {S.~V.}\ \bibnamefont
  {Demishev}}, \bibinfo {author} {\bibfnamefont {A.~V.}\ \bibnamefont
  {Semeno}}, \bibinfo {author} {\bibfnamefont {A.~V.}\ \bibnamefont {Bogach}},
  \bibinfo {author} {\bibfnamefont {V.~V.}\ \bibnamefont {Glushkov}}, \bibinfo
  {author} {\bibfnamefont {N.~E.}\ \bibnamefont {Sluchanko}}, \bibinfo {author}
  {\bibfnamefont {N.~A.}\ \bibnamefont {Samarin}}, \ and\ \bibinfo {author}
  {\bibfnamefont {A.~L.}\ \bibnamefont {Chernobrovkin}},\ }\bibfield  {title}
  {\enquote {\bibinfo {title} {Is \uppercase{M}n\uppercase{S}i an itinerant
  electron magnet? \uppercase{R}esults of \uppercase{ESR} experiments},}\
  }\href {https://doi.org/10.1134/S0021364011040072} {\bibfield  {journal}
  {\bibinfo  {journal} {JETP Letters}\ }\textbf {\bibinfo {volume} {93}},\
  \bibinfo {pages} {213} (\bibinfo {year} {2011})}\BibitemShut {NoStop}%
\bibitem [{\citenamefont {Kittel}(1963)}]{Kittel63}%
  \BibitemOpen
  \bibfield  {author} {\bibinfo {author} {\bibfnamefont {C.}~\bibnamefont
  {Kittel}},\ }\href@noop {} {\emph {\bibinfo {title} {Quantum Theory of
  Solids}}}\ (\bibinfo  {publisher} {J. Wiley \& Sons},\ \bibinfo {address}
  {New York},\ \bibinfo {year} {1963})\BibitemShut {NoStop}%
\bibitem [{\citenamefont {Semadeni}\ \emph {et~al.}(1999)\citenamefont
  {Semadeni}, \citenamefont {B\"oni}, \citenamefont {Endoh}, \citenamefont
  {Roessli},\ and\ \citenamefont {Shirane}}]{Semadeni99}%
  \BibitemOpen
  \bibfield  {author} {\bibinfo {author} {\bibfnamefont {F.}~\bibnamefont
  {Semadeni}}, \bibinfo {author} {\bibfnamefont {P.}~\bibnamefont {B\"oni}},
  \bibinfo {author} {\bibfnamefont {Y.}~\bibnamefont {Endoh}}, \bibinfo
  {author} {\bibfnamefont {B.}~\bibnamefont {Roessli}}, \ and\ \bibinfo
  {author} {\bibfnamefont {G.}~\bibnamefont {Shirane}},\ }\bibfield  {title}
  {\enquote {\bibinfo {title} {Direct observation of spin-flip excitations in
  \uppercase{M}n\uppercase{S}i},}\ }\href
  {https://doi.org/10.1016/S0921-4526(99)00077-0} {\bibfield  {journal}
  {\bibinfo  {journal} {Physica B}\ }\textbf {\bibinfo {volume} {267-268}},\
  \bibinfo {pages} {248} (\bibinfo {year} {1999})}\BibitemShut {NoStop}%
\bibitem [{\citenamefont {Sato}\ \emph {et~al.}(2016)\citenamefont {Sato},
  \citenamefont {Okuyama}, \citenamefont {Hong}, \citenamefont {Kikkawa},
  \citenamefont {Taguchi}, \citenamefont {Arima},\ and\ \citenamefont
  {Tokura}}]{Sato16}%
  \BibitemOpen
  \bibfield  {author} {\bibinfo {author} {\bibfnamefont {Taku~J.}\ \bibnamefont
  {Sato}}, \bibinfo {author} {\bibfnamefont {Daisuke}\ \bibnamefont {Okuyama}},
  \bibinfo {author} {\bibfnamefont {Tao}\ \bibnamefont {Hong}}, \bibinfo
  {author} {\bibfnamefont {Akiko}\ \bibnamefont {Kikkawa}}, \bibinfo {author}
  {\bibfnamefont {Yasujiro}\ \bibnamefont {Taguchi}}, \bibinfo {author}
  {\bibfnamefont {Taka-hisa}\ \bibnamefont {Arima}}, \ and\ \bibinfo {author}
  {\bibfnamefont {Yoshinori}\ \bibnamefont {Tokura}},\ }\bibfield  {title}
  {\enquote {\bibinfo {title} {Magnon dispersion shift in the induced
  ferromagnetic phase of noncentrosymmetric \uppercase{M}n\uppercase{S}i},}\
  }\href {\doibase 10.1103/PhysRevB.94.144420} {\bibfield  {journal} {\bibinfo
  {journal} {Phys. Rev. B}\ }\textbf {\bibinfo {volume} {94}},\ \bibinfo
  {pages} {144420} (\bibinfo {year} {2016})}\BibitemShut {NoStop}%
\bibitem [{\citenamefont {Kugler}\ \emph {et~al.}(2015)\citenamefont {Kugler},
  \citenamefont {Brandl}, \citenamefont {Waizner}, \citenamefont {Janoschek},
  \citenamefont {Georgii}, \citenamefont {Bauer}, \citenamefont {Seemann},
  \citenamefont {Rosch}, \citenamefont {Pfleiderer}, \citenamefont {B\"oni},\
  and\ \citenamefont {Garst}}]{Kugler15}%
  \BibitemOpen
  \bibfield  {author} {\bibinfo {author} {\bibfnamefont {M.}~\bibnamefont
  {Kugler}}, \bibinfo {author} {\bibfnamefont {G.}~\bibnamefont {Brandl}},
  \bibinfo {author} {\bibfnamefont {J.}~\bibnamefont {Waizner}}, \bibinfo
  {author} {\bibfnamefont {M.}~\bibnamefont {Janoschek}}, \bibinfo {author}
  {\bibfnamefont {R.}~\bibnamefont {Georgii}}, \bibinfo {author} {\bibfnamefont
  {A.}~\bibnamefont {Bauer}}, \bibinfo {author} {\bibfnamefont
  {K.}~\bibnamefont {Seemann}}, \bibinfo {author} {\bibfnamefont
  {A.}~\bibnamefont {Rosch}}, \bibinfo {author} {\bibfnamefont
  {C.}~\bibnamefont {Pfleiderer}}, \bibinfo {author} {\bibfnamefont
  {P.}~\bibnamefont {B\"oni}}, \ and\ \bibinfo {author} {\bibfnamefont
  {M.}~\bibnamefont {Garst}},\ }\bibfield  {title} {\enquote {\bibinfo {title}
  {Band structure of helimagnons in \uppercase{M}n\uppercase{S}i resolved by
  inelastic neutron scattering},}\ }\href {\doibase
  10.1103/PhysRevLett.115.097203} {\bibfield  {journal} {\bibinfo  {journal}
  {Phys. Rev. Lett.}\ }\textbf {\bibinfo {volume} {115}},\ \bibinfo {pages}
  {097203} (\bibinfo {year} {2015})}\BibitemShut {NoStop}%
\bibitem [{\citenamefont {Yaouanc}\ \emph {et~al.}(2002)\citenamefont
  {Yaouanc}, \citenamefont {de~R\'eotier}, \citenamefont {Gubbens},
  \citenamefont {Kaiser}, \citenamefont {Menovsky}, \citenamefont {Mihalik},\
  and\ \citenamefont {Cottrell}}]{Yaouanc02}%
  \BibitemOpen
  \bibfield  {author} {\bibinfo {author} {\bibfnamefont {A.}~\bibnamefont
  {Yaouanc}}, \bibinfo {author} {\bibfnamefont {P.~Dalmas}\ \bibnamefont
  {de~R\'eotier}}, \bibinfo {author} {\bibfnamefont {P.~C.~M.}\ \bibnamefont
  {Gubbens}}, \bibinfo {author} {\bibfnamefont {C.~T.}\ \bibnamefont {Kaiser}},
  \bibinfo {author} {\bibfnamefont {A.~A.}\ \bibnamefont {Menovsky}}, \bibinfo
  {author} {\bibfnamefont {M.}~\bibnamefont {Mihalik}}, \ and\ \bibinfo
  {author} {\bibfnamefont {S.~P.}\ \bibnamefont {Cottrell}},\ }\bibfield
  {title} {\enquote {\bibinfo {title} {Evidence for weak itinerant long-range
  magnetic correlations in \uppercase{UG}e$_2$},}\ }\href {\doibase
  10.1103/PhysRevLett.89.147001} {\bibfield  {journal} {\bibinfo  {journal}
  {Phys. Rev. Lett.}\ }\textbf {\bibinfo {volume} {89}},\ \bibinfo {pages}
  {147001} (\bibinfo {year} {2002})}\BibitemShut {NoStop}%
\bibitem [{\citenamefont {Sakarya}\ \emph {et~al.}(2010)\citenamefont
  {Sakarya}, \citenamefont {Gubbens}, \citenamefont {Yaouanc}, \citenamefont
  {{Dalmas de R\'eotier}}, \citenamefont {Andreica}, \citenamefont {Amato},
  \citenamefont {Zimmermann}, \citenamefont {{van Dijk}}, \citenamefont
  {Br\"{u}ck}, \citenamefont {Huang},\ and\ \citenamefont
  {Gortenmulder}}]{Sakarya10}%
  \BibitemOpen
  \bibfield  {author} {\bibinfo {author} {\bibfnamefont {S.}~\bibnamefont
  {Sakarya}}, \bibinfo {author} {\bibfnamefont {P.~C.~M.}\ \bibnamefont
  {Gubbens}}, \bibinfo {author} {\bibfnamefont {A.}~\bibnamefont {Yaouanc}},
  \bibinfo {author} {\bibfnamefont {P.}~\bibnamefont {{Dalmas de R\'eotier}}},
  \bibinfo {author} {\bibfnamefont {D.}~\bibnamefont {Andreica}}, \bibinfo
  {author} {\bibfnamefont {A.}~\bibnamefont {Amato}}, \bibinfo {author}
  {\bibfnamefont {U.}~\bibnamefont {Zimmermann}}, \bibinfo {author}
  {\bibfnamefont {N.~H.}\ \bibnamefont {{van Dijk}}}, \bibinfo {author}
  {\bibfnamefont {E.}~\bibnamefont {Br\"{u}ck}}, \bibinfo {author}
  {\bibfnamefont {Y.}~\bibnamefont {Huang}}, \ and\ \bibinfo {author}
  {\bibfnamefont {T.}~\bibnamefont {Gortenmulder}},\ }\bibfield  {title}
  {\enquote {\bibinfo {title} {Positive muon spin rotation and relaxation
  measurements on the ferromagnetic superconductor \uppercase{UG}e$_2$ at
  ambient and high pressure},}\ }\href@noop {} {\bibfield  {journal} {\bibinfo
  {journal} {Phys. Rev. B}\ }\textbf {\bibinfo {volume} {81}},\ \bibinfo
  {pages} {024429} (\bibinfo {year} {2010})}\BibitemShut {NoStop}%
\bibitem [{Note2()}]{Note2}%
  \BibitemOpen
  \bibinfo {note} {A recent work using a new neutron technique \cite
  {Haslbeck19} confirms the $\mu $SR results.}\BibitemShut {Stop}%
\bibitem [{\citenamefont {Mineev}(2013)}]{Mineev13}%
  \BibitemOpen
  \bibfield  {author} {\bibinfo {author} {\bibfnamefont {V.~P.}\ \bibnamefont
  {Mineev}},\ }\bibfield  {title} {\enquote {\bibinfo {title} {Magnetic
  relaxation in uranium ferromagnetic superconductors},}\ }\href {\doibase
  10.1103/PhysRevB.88.224408} {\bibfield  {journal} {\bibinfo  {journal} {Phys.
  Rev. B}\ }\textbf {\bibinfo {volume} {88}},\ \bibinfo {pages} {224408}
  (\bibinfo {year} {2013})}\BibitemShut {NoStop}%
\bibitem [{\citenamefont {Chubukov}\ \emph {et~al.}(2014)\citenamefont
  {Chubukov}, \citenamefont {Betouras},\ and\ \citenamefont
  {Efremov}}]{Chubukov14}%
  \BibitemOpen
  \bibfield  {author} {\bibinfo {author} {\bibfnamefont {Andrey~V.}\
  \bibnamefont {Chubukov}}, \bibinfo {author} {\bibfnamefont {Joseph~J.}\
  \bibnamefont {Betouras}}, \ and\ \bibinfo {author} {\bibfnamefont
  {Dmitry~V.}\ \bibnamefont {Efremov}},\ }\bibfield  {title} {\enquote
  {\bibinfo {title} {Non-\uppercase{L}andau damping of magnetic excitations in
  systems with localized and itinerant electrons},}\ }\href {\doibase
  10.1103/PhysRevLett.112.037202} {\bibfield  {journal} {\bibinfo  {journal}
  {Phys. Rev. Lett.}\ }\textbf {\bibinfo {volume} {112}},\ \bibinfo {pages}
  {037202} (\bibinfo {year} {2014})}\BibitemShut {NoStop}%
\bibitem [{\citenamefont {Mishra}\ \emph {et~al.}(2016)\citenamefont {Mishra},
  \citenamefont {Krishnan}, \citenamefont {Gangrade}, \citenamefont {Singh},
  \citenamefont {Venkatesh},\ and\ \citenamefont {Ganesan}}]{Mishra16}%
  \BibitemOpen
  \bibfield  {author} {\bibinfo {author} {\bibfnamefont {Ashish}\ \bibnamefont
  {Mishra}}, \bibinfo {author} {\bibfnamefont {M}~\bibnamefont {Krishnan}},
  \bibinfo {author} {\bibfnamefont {Mohan}\ \bibnamefont {Gangrade}}, \bibinfo
  {author} {\bibfnamefont {Durgesh}\ \bibnamefont {Singh}}, \bibinfo {author}
  {\bibfnamefont {R}~\bibnamefont {Venkatesh}}, \ and\ \bibinfo {author}
  {\bibfnamefont {V}~\bibnamefont {Ganesan}},\ }\bibfield  {title} {\enquote
  {\bibinfo {title} {Low temperature heat capacity of polycrystalline
  \uppercase{M}n\uppercase{S}i},}\ }\href
  {https://doi.org/10.1088/1742-6596/755/1/012037} {\bibfield  {journal}
  {\bibinfo  {journal} {J. Phys.: Conf. Ser.}\ }\textbf {\bibinfo {volume}
  {755}},\ \bibinfo {pages} {012037} (\bibinfo {year} {2016})}\BibitemShut
  {NoStop}%
\bibitem [{\citenamefont {Bos}\ \emph {et~al.}(2008)\citenamefont {Bos},
  \citenamefont {Colin},\ and\ \citenamefont {Palstra}}]{Bos08}%
  \BibitemOpen
  \bibfield  {author} {\bibinfo {author} {\bibfnamefont {Jan-Willem~G.}\
  \bibnamefont {Bos}}, \bibinfo {author} {\bibfnamefont {Claire~V.}\
  \bibnamefont {Colin}}, \ and\ \bibinfo {author} {\bibfnamefont {Thomas
  T.~M.}\ \bibnamefont {Palstra}},\ }\bibfield  {title} {\enquote {\bibinfo
  {title} {Magnetoelectric coupling in the cubic ferrimagnet
  \uppercase{C}u$_2$\uppercase{OS}e\uppercase{O}$_3$},}\ }\href {\doibase
  10.1103/PhysRevB.78.094416} {\bibfield  {journal} {\bibinfo  {journal} {Phys.
  Rev. B}\ }\textbf {\bibinfo {volume} {78}},\ \bibinfo {pages} {094416}
  (\bibinfo {year} {2008})}\BibitemShut {NoStop}%
\bibitem [{\citenamefont {Maisuradze}\ \emph {et~al.}(2011)\citenamefont
  {Maisuradze}, \citenamefont {Guguchia}, \citenamefont {Graneli},
  \citenamefont {R\o{}nnow}, \citenamefont {Berger},\ and\ \citenamefont
  {Keller}}]{Maisuradze11}%
  \BibitemOpen
  \bibfield  {author} {\bibinfo {author} {\bibfnamefont {A.}~\bibnamefont
  {Maisuradze}}, \bibinfo {author} {\bibfnamefont {Z.}~\bibnamefont
  {Guguchia}}, \bibinfo {author} {\bibfnamefont {B.}~\bibnamefont {Graneli}},
  \bibinfo {author} {\bibfnamefont {H.~M.}\ \bibnamefont {R\o{}nnow}}, \bibinfo
  {author} {\bibfnamefont {H.}~\bibnamefont {Berger}}, \ and\ \bibinfo {author}
  {\bibfnamefont {H.}~\bibnamefont {Keller}},\ }\bibfield  {title} {\enquote
  {\bibinfo {title} {$\ensuremath{\mu}$\uppercase{SR} investigation of
  magnetism and magnetoelectric coupling in
  \uppercase{C}u${}_{2}$\uppercase{OS}e\uppercase{O}${}_{3}$},}\ }\href
  {\doibase 10.1103/PhysRevB.84.064433} {\bibfield  {journal} {\bibinfo
  {journal} {Phys. Rev. B}\ }\textbf {\bibinfo {volume} {84}},\ \bibinfo
  {pages} {064433} (\bibinfo {year} {2011})}\BibitemShut {NoStop}%
\bibitem [{\citenamefont {Adams}\ \emph {et~al.}(2012)\citenamefont {Adams},
  \citenamefont {Chacon}, \citenamefont {Wagner}, \citenamefont {Bauer},
  \citenamefont {Brandl}, \citenamefont {Pedersen}, \citenamefont {Berger},
  \citenamefont {Lemmens},\ and\ \citenamefont {Pfleiderer}}]{Adams12}%
  \BibitemOpen
  \bibfield  {author} {\bibinfo {author} {\bibfnamefont {T.}~\bibnamefont
  {Adams}}, \bibinfo {author} {\bibfnamefont {A.}~\bibnamefont {Chacon}},
  \bibinfo {author} {\bibfnamefont {M.}~\bibnamefont {Wagner}}, \bibinfo
  {author} {\bibfnamefont {A.}~\bibnamefont {Bauer}}, \bibinfo {author}
  {\bibfnamefont {G.}~\bibnamefont {Brandl}}, \bibinfo {author} {\bibfnamefont
  {B.}~\bibnamefont {Pedersen}}, \bibinfo {author} {\bibfnamefont
  {H.}~\bibnamefont {Berger}}, \bibinfo {author} {\bibfnamefont
  {P.}~\bibnamefont {Lemmens}}, \ and\ \bibinfo {author} {\bibfnamefont
  {C.}~\bibnamefont {Pfleiderer}},\ }\bibfield  {title} {\enquote {\bibinfo
  {title} {Long-wavelength helimagnetic order and skyrmion lattice phase in
  \uppercase{C}u$_2$\uppercase{OS}e\uppercase{O}$_3$},}\ }\href {\doibase
  10.1103/PhysRevLett.108.237204} {\bibfield  {journal} {\bibinfo  {journal}
  {Phys. Rev. Lett.}\ }\textbf {\bibinfo {volume} {108}},\ \bibinfo {pages}
  {237204} (\bibinfo {year} {2012})}\BibitemShut {NoStop}%
\bibitem [{\citenamefont {Choi}\ \emph {et~al.}(2019)\citenamefont {Choi},
  \citenamefont {Tai},\ and\ \citenamefont {Zhu}}]{Choi19}%
  \BibitemOpen
  \bibfield  {author} {\bibinfo {author} {\bibfnamefont {Hongchul}\
  \bibnamefont {Choi}}, \bibinfo {author} {\bibfnamefont {Yuan-Yen}\
  \bibnamefont {Tai}}, \ and\ \bibinfo {author} {\bibfnamefont {Jian-Xin}\
  \bibnamefont {Zhu}},\ }\bibfield  {title} {\enquote {\bibinfo {title}
  {Spin-fermion model for skyrmions in \uppercase{M}n\uppercase{G}e derived
  from strong correlations},}\ }\href {\doibase 10.1103/PhysRevB.99.134437}
  {\bibfield  {journal} {\bibinfo  {journal} {Phys. Rev. B}\ }\textbf {\bibinfo
  {volume} {99}},\ \bibinfo {pages} {134437} (\bibinfo {year}
  {2019})}\BibitemShut {NoStop}%
\bibitem [{\citenamefont {Xiang~Chen}\ \emph {et~al.}(2019)\citenamefont
  {Xiang~Chen}, \citenamefont {Stone}, \citenamefont {Kolesnikov},
  \citenamefont {Wolf}, \citenamefont {Reznik}, \citenamefont {Bedell},
  \citenamefont {Lechermann},\ and\ \citenamefont {Wilson}}]{Chen19}%
  \BibitemOpen
  \bibfield  {author} {\bibinfo {author} {\bibfnamefont {Igor~Krivenko}\
  \bibnamefont {Xiang~Chen}}, \bibinfo {author} {\bibfnamefont {Matthew~B.}\
  \bibnamefont {Stone}}, \bibinfo {author} {\bibfnamefont {Alexander~I.}\
  \bibnamefont {Kolesnikov}}, \bibinfo {author} {\bibfnamefont {Thomas}\
  \bibnamefont {Wolf}}, \bibinfo {author} {\bibfnamefont {Dmitry}\ \bibnamefont
  {Reznik}}, \bibinfo {author} {\bibfnamefont {Kevin~S.}\ \bibnamefont
  {Bedell}}, \bibinfo {author} {\bibfnamefont {Frank}\ \bibnamefont
  {Lechermann}}, \ and\ \bibinfo {author} {\bibfnamefont {Stephen~D.}\
  \bibnamefont {Wilson}},\ }\href@noop {} {\enquote {\bibinfo {title}
  {Unconventional \uppercase{H}und metal in \uppercase{M}n\uppercase{S}i},}\ }
  (\bibinfo {year} {2019}),\ \bibinfo {note} {arXiv:1909.11195}\BibitemShut
  {NoStop}%
\bibitem [{\citenamefont {Dalmas~de R\'eotier}\ and\ \citenamefont
  {Yaouanc}(1995)}]{Dalmas95}%
  \BibitemOpen
  \bibfield  {author} {\bibinfo {author} {\bibfnamefont {P.}~\bibnamefont
  {Dalmas~de R\'eotier}}\ and\ \bibinfo {author} {\bibfnamefont
  {A.}~\bibnamefont {Yaouanc}},\ }\bibfield  {title} {\enquote {\bibinfo
  {title} {Zero-field muon spin lattice relaxation rate in a
  \uppercase{H}eisenberg ferromagnet at low temperature},}\ }\href {\doibase
  10.1103/PhysRevB.52.9155} {\bibfield  {journal} {\bibinfo  {journal} {Phys.
  Rev. B}\ }\textbf {\bibinfo {volume} {52}},\ \bibinfo {pages} {9155--9158}
  (\bibinfo {year} {1995})}\BibitemShut {NoStop}%
\bibitem [{Note3()}]{Note3}%
  \BibitemOpen
  \bibinfo {note} {The muon spin relaxation rate associated to magnon Raman
  scattering is estimated to be $10^{-5}$~$\mu $s$^{-1}$ at 10~K; see Eq.
  10.218 in Ref.~\protect \rev@citealp {Yaouanc11}. The spin wave stiffness
  constant $D_{\protect \rm FM}$ therein is directly related to $c_\perp
  $.}\BibitemShut {Stop}%
\bibitem [{\citenamefont {Moriya}(1956{\natexlab{a}})}]{Moriya56a}%
  \BibitemOpen
  \bibfield  {author} {\bibinfo {author} {\bibfnamefont {T.}~\bibnamefont
  {Moriya}},\ }\bibfield  {title} {\enquote {\bibinfo {title} {Nuclear magnetic
  relaxation in antiferromagnetics},}\ }\href@noop {} {\bibfield  {journal}
  {\bibinfo  {journal} {Prog. Theor. Phys. (Kyoto)}\ }\textbf {\bibinfo
  {volume} {16}},\ \bibinfo {pages} {23--44} (\bibinfo {year}
  {1956}{\natexlab{a}})}\BibitemShut {NoStop}%
\bibitem [{\citenamefont {Moriya}(1956{\natexlab{b}})}]{Moriya56b}%
  \BibitemOpen
  \bibfield  {author} {\bibinfo {author} {\bibfnamefont {T.}~\bibnamefont
  {Moriya}},\ }\bibfield  {title} {\enquote {\bibinfo {title} {Nuclear magnetic
  relaxation in antiferromagnetics, \uppercase{II}},}\ }\href@noop {}
  {\bibfield  {journal} {\bibinfo  {journal} {Prog. Theor. Phys. (Kyoto)}\
  }\textbf {\bibinfo {volume} {16}},\ \bibinfo {pages} {641--657} (\bibinfo
  {year} {1956}{\natexlab{b}})}\BibitemShut {NoStop}%
\bibitem [{\citenamefont {van Kranendonk}\ and\ \citenamefont
  {Bloom}(1956)}]{vanKranendonk56}%
  \BibitemOpen
  \bibfield  {author} {\bibinfo {author} {\bibfnamefont {J.}~\bibnamefont {van
  Kranendonk}}\ and\ \bibinfo {author} {\bibfnamefont {M.}~\bibnamefont
  {Bloom}},\ }\bibfield  {title} {\enquote {\bibinfo {title} {Nuclear
  relaxation in antiferromagnetic crystals},}\ }\href@noop {} {\bibfield
  {journal} {\bibinfo  {journal} {Physica}\ }\textbf {\bibinfo {volume} {22}},\
  \bibinfo {pages} {545} (\bibinfo {year} {1956})}\BibitemShut {NoStop}%
\bibitem [{\citenamefont {Mitchell}(1957)}]{Mitchell57}%
  \BibitemOpen
  \bibfield  {author} {\bibinfo {author} {\bibfnamefont {A.~H.}\ \bibnamefont
  {Mitchell}},\ }\bibfield  {title} {\enquote {\bibinfo {title} {Nuclear
  relaxation by the hyperfine interaction with the ion core spins in
  ferromagnetic and antiferromagnetic crystals},}\ }\href@noop {} {\bibfield
  {journal} {\bibinfo  {journal} {J. Chem. Phys.}\ }\textbf {\bibinfo {volume}
  {27}},\ \bibinfo {pages} {17--19} (\bibinfo {year} {1957})}\BibitemShut
  {NoStop}%
\bibitem [{\citenamefont {Beeman}\ and\ \citenamefont
  {Pincus}(1968)}]{Beeman68}%
  \BibitemOpen
  \bibfield  {author} {\bibinfo {author} {\bibfnamefont {D.}~\bibnamefont
  {Beeman}}\ and\ \bibinfo {author} {\bibfnamefont {P.}~\bibnamefont
  {Pincus}},\ }\bibfield  {title} {\enquote {\bibinfo {title} {Nuclear
  spin-lattice relaxation in magnetic insulators},}\ }\href {\doibase
  10.1103/PhysRev.166.359} {\bibfield  {journal} {\bibinfo  {journal} {Phys.
  Rev.}\ }\textbf {\bibinfo {volume} {166}},\ \bibinfo {pages} {359--375}
  (\bibinfo {year} {1968})}\BibitemShut {NoStop}%
\bibitem [{\citenamefont {Yosida}\ and\ \citenamefont {Miwa}(1961)}]{Yosida61}%
  \BibitemOpen
  \bibfield  {author} {\bibinfo {author} {\bibfnamefont {K.}~\bibnamefont
  {Yosida}}\ and\ \bibinfo {author} {\bibfnamefont {H.}~\bibnamefont {Miwa}},\
  }\bibfield  {title} {\enquote {\bibinfo {title} {Magnetic ordering in the
  ferromagnetic rare-earth metals},}\ }\href {\doibase
  https://doi.org/10.1063/1.2000511} {\bibfield  {journal} {\bibinfo  {journal}
  {J. Appl. Phys.}\ }\textbf {\bibinfo {volume} {32}},\ \bibinfo {pages} {8S}
  (\bibinfo {year} {1961})}\BibitemShut {NoStop}%
\bibitem [{\citenamefont {Cooper}\ \emph {et~al.}(1962)\citenamefont {Cooper},
  \citenamefont {Elliott}, \citenamefont {Nettel},\ and\ \citenamefont
  {Suhl}}]{Cooper62}%
  \BibitemOpen
  \bibfield  {author} {\bibinfo {author} {\bibfnamefont {B.~R.}\ \bibnamefont
  {Cooper}}, \bibinfo {author} {\bibfnamefont {R.~J.}\ \bibnamefont {Elliott}},
  \bibinfo {author} {\bibfnamefont {S.~J.}\ \bibnamefont {Nettel}}, \ and\
  \bibinfo {author} {\bibfnamefont {H.}~\bibnamefont {Suhl}},\ }\bibfield
  {title} {\enquote {\bibinfo {title} {Theory of magnetic resonance in the
  heavy rare-earth metals},}\ }\href {\doibase 10.1103/PhysRev.127.57}
  {\bibfield  {journal} {\bibinfo  {journal} {Phys. Rev.}\ }\textbf {\bibinfo
  {volume} {127}},\ \bibinfo {pages} {57--68} (\bibinfo {year}
  {1962})}\BibitemShut {NoStop}%
\bibitem [{\citenamefont {Izyumov}\ and\ \citenamefont
  {Laptev}(1985)}]{Izyumov85}%
  \BibitemOpen
  \bibfield  {author} {\bibinfo {author} {\bibfnamefont {Yu.~A}\ \bibnamefont
  {Izyumov}}\ and\ \bibinfo {author} {\bibfnamefont {V.~M.}\ \bibnamefont
  {Laptev}},\ }\bibfield  {title} {\enquote {\bibinfo {title} {Excitation
  spectrum of incommensurable magnetic structures and neutron scattering},}\
  }\href@noop {} {\bibfield  {journal} {\bibinfo  {journal} {Sov. Phys. JETP}\
  }\textbf {\bibinfo {volume} {61}},\ \bibinfo {pages} {95} (\bibinfo {year}
  {1985})}\BibitemShut {NoStop}%
\bibitem [{\citenamefont {Overhauser}(1971)}]{Overhauser71}%
  \BibitemOpen
  \bibfield  {author} {\bibinfo {author} {\bibfnamefont {A.~W.}\ \bibnamefont
  {Overhauser}},\ }\bibfield  {title} {\enquote {\bibinfo {title}
  {Observability of charge-density waves by neutron diffraction},}\ }\href
  {\doibase 10.1103/PhysRevB.3.3173} {\bibfield  {journal} {\bibinfo  {journal}
  {Phys. Rev. B}\ }\textbf {\bibinfo {volume} {3}},\ \bibinfo {pages} {3173}
  (\bibinfo {year} {1971})}\BibitemShut {NoStop}%
\bibitem [{\citenamefont {Yaouanc}\ \emph
  {et~al.}(1993{\natexlab{a}})\citenamefont {Yaouanc}, \citenamefont {{Dalmas
  de R\'eotier}},\ and\ \citenamefont {Frey}}]{Yaouanc93}%
  \BibitemOpen
  \bibfield  {author} {\bibinfo {author} {\bibfnamefont {A.}~\bibnamefont
  {Yaouanc}}, \bibinfo {author} {\bibfnamefont {P.}~\bibnamefont {{Dalmas de
  R\'eotier}}}, \ and\ \bibinfo {author} {\bibfnamefont {E.}~\bibnamefont
  {Frey}},\ }\bibfield  {title} {\enquote {\bibinfo {title} {Zero-field
  muon-spin-relaxation depolarization rate of paramagnets near the
  \uppercase{C}urie temperature},}\ }\href {\doibase 10.1103/PhysRevB.47.796}
  {\bibfield  {journal} {\bibinfo  {journal} {Phys. Rev. B}\ }\textbf {\bibinfo
  {volume} {47}},\ \bibinfo {pages} {796} (\bibinfo {year}
  {1993}{\natexlab{a}})}\BibitemShut {NoStop}%
\bibitem [{\citenamefont {Yaouanc}\ \emph
  {et~al.}(1993{\natexlab{b}})\citenamefont {Yaouanc}, \citenamefont {{Dalmas
  de R\'eotier}},\ and\ \citenamefont {Frey}}]{Yaouanc93a}%
  \BibitemOpen
  \bibfield  {author} {\bibinfo {author} {\bibfnamefont {A.}~\bibnamefont
  {Yaouanc}}, \bibinfo {author} {\bibfnamefont {P.}~\bibnamefont {{Dalmas de
  R\'eotier}}}, \ and\ \bibinfo {author} {\bibfnamefont {E.}~\bibnamefont
  {Frey}},\ }\bibfield  {title} {\enquote {\bibinfo {title} {Probing
  longitudinal and transverse spin dynamics of paramagnets near ${T}_{\rm {c}}$
  by zero-field $\mu$\uppercase{SR} measurements},}\ }\href
  {http://stacks.iop.org/0295-5075/21/i=1/a=016} {\bibfield  {journal}
  {\bibinfo  {journal} {Europhys. Lett.}\ }\textbf {\bibinfo {volume} {21}},\
  \bibinfo {pages} {93} (\bibinfo {year} {1993}{\natexlab{b}})}\BibitemShut
  {NoStop}%
\bibitem [{\citenamefont {{Dalmas de R\'eotier}}\ and\ \citenamefont
  {Yaouanc}(1994)}]{Dalmas94}%
  \BibitemOpen
  \bibfield  {author} {\bibinfo {author} {\bibfnamefont {P.}~\bibnamefont
  {{Dalmas de R\'eotier}}}\ and\ \bibinfo {author} {\bibfnamefont
  {A.}~\bibnamefont {Yaouanc}},\ }\bibfield  {title} {\enquote {\bibinfo
  {title} {Possibility of observation of the critical paramagnetic longitudinal
  spin fluctuations in \uppercase{G}adolinium by muon spin rotation
  spectroscopy},}\ }\href {\doibase 10.1103/PhysRevLett.72.290} {\bibfield
  {journal} {\bibinfo  {journal} {Phys. Rev. Lett.}\ }\textbf {\bibinfo
  {volume} {72}},\ \bibinfo {pages} {290--293} (\bibinfo {year}
  {1994})}\BibitemShut {NoStop}%
\bibitem [{\citenamefont {Frey}\ \emph {et~al.}(1997)\citenamefont {Frey},
  \citenamefont {Schwabl}, \citenamefont {Henneberger}, \citenamefont
  {Hartmann}, \citenamefont {W\"appling}, \citenamefont {Kratzer},\ and\
  \citenamefont {Kalvius}}]{Frey97}%
  \BibitemOpen
  \bibfield  {author} {\bibinfo {author} {\bibfnamefont {E.}~\bibnamefont
  {Frey}}, \bibinfo {author} {\bibfnamefont {F.}~\bibnamefont {Schwabl}},
  \bibinfo {author} {\bibfnamefont {S.}~\bibnamefont {Henneberger}}, \bibinfo
  {author} {\bibfnamefont {O.}~\bibnamefont {Hartmann}}, \bibinfo {author}
  {\bibfnamefont {R.}~\bibnamefont {W\"appling}}, \bibinfo {author}
  {\bibfnamefont {A.}~\bibnamefont {Kratzer}}, \ and\ \bibinfo {author}
  {\bibfnamefont {G.~M.}\ \bibnamefont {Kalvius}},\ }\bibfield  {title}
  {\enquote {\bibinfo {title} {Determination of the universality class of
  \uppercase{G}adolinium},}\ }\href {\doibase 10.1103/PhysRevLett.79.5142}
  {\bibfield  {journal} {\bibinfo  {journal} {Phys. Rev. Lett.}\ }\textbf
  {\bibinfo {volume} {79}},\ \bibinfo {pages} {5142--5145} (\bibinfo {year}
  {1997})}\BibitemShut {NoStop}%
\bibitem [{\citenamefont {Yaouanc}\ \emph {et~al.}(1996)\citenamefont
  {Yaouanc}, \citenamefont {{Dalmas de R\'eotier}}, \citenamefont {Gubbens},
  \citenamefont {Mulders}, \citenamefont {Kayzel},\ and\ \citenamefont
  {Franse}}]{Yaouanc96}%
  \BibitemOpen
  \bibfield  {author} {\bibinfo {author} {\bibfnamefont {A.}~\bibnamefont
  {Yaouanc}}, \bibinfo {author} {\bibfnamefont {P.}~\bibnamefont {{Dalmas de
  R\'eotier}}}, \bibinfo {author} {\bibfnamefont {P.C.M.}\ \bibnamefont
  {Gubbens}}, \bibinfo {author} {\bibfnamefont {A.~M.}\ \bibnamefont
  {Mulders}}, \bibinfo {author} {\bibfnamefont {F.~E.}\ \bibnamefont {Kayzel}},
  \ and\ \bibinfo {author} {\bibfnamefont {J.~J.~M.}\ \bibnamefont {Franse}},\
  }\bibfield  {title} {\enquote {\bibinfo {title} {Muon-spin-relaxation study
  of the critical longitudinal spin dynamics in a dipolar
  \uppercase{H}eisenberg ferromagnet},}\ }\href@noop {} {\bibfield  {journal}
  {\bibinfo  {journal} {Phys. Rev. B}\ }\textbf {\bibinfo {volume} {53}},\
  \bibinfo {pages} {350} (\bibinfo {year} {1996})}\BibitemShut {NoStop}%
\bibitem [{Note4()}]{Note4}%
  \BibitemOpen
  \bibinfo {note} {The notation for $R(\varphi ,\theta ,\psi )$ follows that of
  Eq.~D6 in Ref.~\protect \rev@citealp {Yaouanc11}}\BibitemShut {NoStop}%
\bibitem [{\citenamefont {Grigoriev}\ \emph {et~al.}(2006)\citenamefont
  {Grigoriev}, \citenamefont {Maleyev}, \citenamefont {Okorokov}, \citenamefont
  {Chetverikov}, \citenamefont {B\"oni}, \citenamefont {Georgii}, \citenamefont
  {Lamago}, \citenamefont {Eckerlebe},\ and\ \citenamefont
  {Pranzas}}]{Grigoriev06}%
  \BibitemOpen
  \bibfield  {author} {\bibinfo {author} {\bibfnamefont {S.~V.}\ \bibnamefont
  {Grigoriev}}, \bibinfo {author} {\bibfnamefont {S.~V.}\ \bibnamefont
  {Maleyev}}, \bibinfo {author} {\bibfnamefont {A.~I.}\ \bibnamefont
  {Okorokov}}, \bibinfo {author} {\bibfnamefont {Yu.~O.}\ \bibnamefont
  {Chetverikov}}, \bibinfo {author} {\bibfnamefont {P.}~\bibnamefont {B\"oni}},
  \bibinfo {author} {\bibfnamefont {R.}~\bibnamefont {Georgii}}, \bibinfo
  {author} {\bibfnamefont {D.}~\bibnamefont {Lamago}}, \bibinfo {author}
  {\bibfnamefont {H.}~\bibnamefont {Eckerlebe}}, \ and\ \bibinfo {author}
  {\bibfnamefont {K.}~\bibnamefont {Pranzas}},\ }\bibfield  {title} {\enquote
  {\bibinfo {title} {Magnetic structure of \uppercase{M}n\uppercase{S}i under
  an applied field probed by polarized small-angle neutron scattering},}\
  }\href {\doibase 10.1103/PhysRevB.74.214414} {\bibfield  {journal} {\bibinfo
  {journal} {Phys. Rev. B}\ }\textbf {\bibinfo {volume} {74}},\ \bibinfo
  {pages} {214414} (\bibinfo {year} {2006})}\BibitemShut {NoStop}%
\bibitem [{\citenamefont {Janoschek}\ \emph {et~al.}(2013)\citenamefont
  {Janoschek}, \citenamefont {Garst}, \citenamefont {Bauer}, \citenamefont
  {Krautscheid}, \citenamefont {Georgii}, \citenamefont {B\"oni},\ and\
  \citenamefont {Pfleiderer}}]{Janoschek13}%
  \BibitemOpen
  \bibfield  {author} {\bibinfo {author} {\bibfnamefont {M.}~\bibnamefont
  {Janoschek}}, \bibinfo {author} {\bibfnamefont {M.}~\bibnamefont {Garst}},
  \bibinfo {author} {\bibfnamefont {A.}~\bibnamefont {Bauer}}, \bibinfo
  {author} {\bibfnamefont {P.}~\bibnamefont {Krautscheid}}, \bibinfo {author}
  {\bibfnamefont {R.}~\bibnamefont {Georgii}}, \bibinfo {author} {\bibfnamefont
  {P.}~\bibnamefont {B\"oni}}, \ and\ \bibinfo {author} {\bibfnamefont
  {C.}~\bibnamefont {Pfleiderer}},\ }\bibfield  {title} {\enquote {\bibinfo
  {title} {Fluctuation-induced first-order phase transition in
  \uppercase{D}zyaloshinskii-\uppercase{M}oriya helimagnets},}\ }\href
  {\doibase 10.1103/PhysRevB.87.134407} {\bibfield  {journal} {\bibinfo
  {journal} {Phys. Rev. B}\ }\textbf {\bibinfo {volume} {87}},\ \bibinfo
  {pages} {134407} (\bibinfo {year} {2013})}\BibitemShut {NoStop}%
\bibitem [{\citenamefont {Haslbeck}\ \emph {et~al.}(2019)\citenamefont
  {Haslbeck}, \citenamefont {S\"aubert}, \citenamefont {Seifert}, \citenamefont
  {Franz}, \citenamefont {Schulz}, \citenamefont {Heinemann}, \citenamefont
  {Keller}, \citenamefont {Das}, \citenamefont {Thompson}, \citenamefont
  {Bauer}, \citenamefont {Pfleiderer},\ and\ \citenamefont
  {Janoschek}}]{Haslbeck19}%
  \BibitemOpen
  \bibfield  {author} {\bibinfo {author} {\bibfnamefont {F.}~\bibnamefont
  {Haslbeck}}, \bibinfo {author} {\bibfnamefont {S.}~\bibnamefont {S\"aubert}},
  \bibinfo {author} {\bibfnamefont {M.}~\bibnamefont {Seifert}}, \bibinfo
  {author} {\bibfnamefont {C.}~\bibnamefont {Franz}}, \bibinfo {author}
  {\bibfnamefont {M.}~\bibnamefont {Schulz}}, \bibinfo {author} {\bibfnamefont
  {A.}~\bibnamefont {Heinemann}}, \bibinfo {author} {\bibfnamefont
  {T.}~\bibnamefont {Keller}}, \bibinfo {author} {\bibfnamefont {Pinaki}\
  \bibnamefont {Das}}, \bibinfo {author} {\bibfnamefont {J.~D.}\ \bibnamefont
  {Thompson}}, \bibinfo {author} {\bibfnamefont {E.~D.}\ \bibnamefont {Bauer}},
  \bibinfo {author} {\bibfnamefont {C.}~\bibnamefont {Pfleiderer}}, \ and\
  \bibinfo {author} {\bibfnamefont {M.}~\bibnamefont {Janoschek}},\ }\bibfield
  {title} {\enquote {\bibinfo {title} {Ultrahigh-resolution neutron
  spectroscopy of low-energy spin dynamics in \uppercase{UG}e$_{2}$},}\ }\href
  {\doibase 10.1103/PhysRevB.99.014429} {\bibfield  {journal} {\bibinfo
  {journal} {Phys. Rev. B}\ }\textbf {\bibinfo {volume} {99}},\ \bibinfo
  {pages} {014429} (\bibinfo {year} {2019})}\BibitemShut {NoStop}%
\end{thebibliography}%

\end{document}